\documentclass[%
 reprint,
 amsmath,amssymb,
 aps,
 pr,
 longbibliography,
floatfix,
showkeys,
]{revtex4-1}

\usepackage{graphicx}
\usepackage{dcolumn}
\usepackage{bm}
\DeclareMathOperator{\sinc}{sinc}
\usepackage{array}
\usepackage{tabularray}
\usepackage{xcolor}

\AtBeginDocument{%
    \newwrite\bibnotes
    \def\bibnotesext{refs.bib}
    \immediate\openout\bibnotes=\jobname\bibnotesext
    \immediate\write\bibnotes{@CONTROL{REVTEX41Control}}
    \immediate\write\bibnotes{@CONTROL{%
    apsrev41Control,author="08",editor="1",pages="1",title="0",year="1"}}
     \if@filesw
     \immediate\write\@auxout{\string\citation{apsrev41Control}}%
    \fi
}

\begin{document}

\preprint{APS/123-QED}

\title{Entangling entanglement: coupling frequency and polarization of biphotons on demand}

\author{Arash Riazi}
\affiliation{
Department of Electrical and Computer Engineering, University of Toronto, Toronto, Ontario, M5S 3G4, Canada
}
\author{Eric Y. Zhu}
\affiliation{
Department of Electrical and Computer Engineering, University of Toronto, Toronto, Ontario, M5S 3G4, Canada
}
\author{Dan Xu}%
 \email{xudan.xu@utoronto.ca}
\affiliation{
Department of Electrical and Computer Engineering, University of Toronto, Toronto, Ontario, M5S 3G4, Canada
}

\author{Li Qian}
 \email{l.qian@utoronto.ca}
\affiliation{
Department of Electrical and Computer Engineering, University of Toronto, Toronto, Ontario, M5S 3G4, Canada
}%

\date{\today}

\begin{abstract}
Quantum information is often carried in the frequency and polarization degrees of freedom (DoFs) in single photons and entangled photons. We demonstrate a new approach to couple and decouple the frequency and polarization DoFs of broadband biphotons. Our approach is based on a common-path nonlinear interferometer (CP-NLI) with a linear dispersive medium and a polarization controller sandwiched in between two nonlinear media that generate the interfering biphotons. By adjusting the polarization controller, we can effectively manipulate the two DoFs. When the two DoFs are decoupled, maximally polarization-entangled biphotons are observed in the polarization DoF, while interference fringes are observed in the spectral intensity of the biphotons. When the two DoFs are coupled, however, interference fringes disappear from the spectral intensity and instead appear in the degree of polarization entanglement. The degree of polarization entanglement quantified by concurrence in principle can vary from 0 to 1 depending on the signal and idler photon frequencies. Our approach offers a convenient means of tuning the polarization entanglement and can be employed for arbitrary biphoton polarization state generation, with applications in quantum information processing and the study of fundamental physics.
\end{abstract}

\maketitle

\section{\label{Intro}Introduction}
Much of the development in the fields of quantum optics and quantum information processing relies on our ability to effectively manipulate biphoton states in various degrees of freedom (DoFs). Biphoton manipulation is in fact a necessary step towards generating arbitrary biphoton states required for many quantum applications\,\cite{jennewein_quantum_2000,kues_-chip_2017,riazi_biphoton_2019,williams_quantum_2019,vallone_realization_2007,chen_experimental_2007} and for studying fundamental physics\,\cite{shalm_strong_2015,wei_synthesizing_2005}.

In the realm of generating biphotons, nonlinear optical processes, such as spontaneous parametric down-conversion (SPDC), have traditionally served as workhorses\,\cite{brendel_pulsed_1999,mair_entanglement_2001,kwiat_new_1995}. Nevertheless,  they are usually incapable of offering a dynamic way of manipulating the properties of the generated biphotons in various DoFs. Hence, additional steps are required either before or after the generation of biphotons from a nonlinear process\,\cite{wei_synthesizing_2005,thew_experimental_2002,bernhard_shaping_2013,kwiat_ultrabright_1999,mattle_dense_1996,wu_revival_2019,barreiro_generation_2005,chen_recovering_2020} to \textit{directly} manipulate biphotons in various DoFs.

Direct manipulation of biphotons within a specific DoF, however, may not always be straightforward. An alternative approach has been proposed, involving the \textit{indirect} manipulation of biphotons through the utilization of an ancillary DoF to establish \textit{coupling} with the target DoF\,\cite{guo_frequency-bin_2015,shu_narrowband_2015,chen_frequency-induced_2015}. Manipulation of the more accessible ancillary DoF will result in an appropriate change of the target DoF of biphotons. This form of \textit{indirect} biphoton manipulation requires a reliable way of coupling between DoFs of biphoton state in the first place. Several techniques have been reported in the literature to facilitate such coupling, including the use of linear optics and mode selection\,\cite{vallone_realization_2007}, as well as the incorporation of active elements such as frequency shifters\,\cite{guo_frequency-bin_2015,shu_narrowband_2015}. 

Even though biphotons with coupled DoFs have a unique role in generating complex photonics states, there are few studies offering techniques for generating such states, including cluster states\,\cite{vallone_realization_2007,ciampini_path-polarization_2016,reimer_high-dimensional_2019}. In contrast, within the classical realm, there is a well-established practice of coupling various DoFs to create classical nonseparable states of light, often referred to as 'classical entanglement'\,\cite{spreeuw_classical_1998,karimi_classical_2015,shen_nonseparable_2022}. This is commonly achieved by coupling the polarization and spatial modes of the electromagnetic field through the use of spatial light modulators, liquid crystal \textit{q}-plates, \textit{J}-plates, or metamaterials\,\cite{shen_creation_2021,marrucci_optical_2006,devlin_arbitrary_2017,wang_multichannel_2019}. Nonetheless, it's crucial to note that this form of coupling does not offer the capability to independently control one DoF by adjusting the other.

Here, we propose and demonstrate a new method for generating biphotons with coupled DoFs (polarization and frequency), by utilizing a common-path nonlinear interferometer (CP-NLI) without the need for postselection or active elements. The method allows for the coupling and decoupling of the two DoFs of biphotons on demand through a simple polarization transformation. As a result of the coupling between two DoFs, biphoton properties in polarization DoF (such as polarization entanglement) can be controlled and tuned by choosing different frequencies of biphotons.   

Prior to providing a comprehensive explanation of our technique, we would like to remark that biphotons with coupled DoFs differ from two other commonly known categories of biphotons, namely hybrid- and hyper-entangled biphotons. Hyper-entangled biphoton states\,\cite{barreiro_generation_2005} can be expressed as:
\begin{equation}
|\psi_\text{hyper}\rangle=|\psi^\text{DoF 1}\rangle\otimes|\psi^\text{DoF 2}\rangle\otimes\cdots\otimes|\psi^{\text{DoF}\,n}\rangle,
\label{eq1}
\end{equation}
where $|\psi^{\text{DoF}\,n}\rangle$ is the maximally-entangled biphoton quantum state in DoF $n$. The hyper-entangled biphoton state consists of a tensor product in each DoF ($|\psi^{\text{DoF}\,n}\rangle$) and there exists no coupling between these DoFs. 

In the case of hybrid-entangled biphotons\,\cite{neves_hybrid_2009,ma_experimental_2009,gabriel_entangling_2011}, one DoF of one photon is entangled to a different DoF of the other photon in the pair. The most simple form of such a biphoton state can be written as:
\begin{equation}
|\psi_\text{hybrid}\rangle=\frac{1}{\sqrt{2}}(|\zeta_1\rangle\otimes|\eta_2\rangle+|\zeta_1^\perp\rangle\otimes|\eta_2^\perp\rangle),
\label{eq2}
\end{equation}
where $|\zeta_1\rangle(|\zeta_1^\perp\rangle)$ and $|\eta_2\rangle(|\eta_2^\perp\rangle)$ are the orthonormal bases in the Hilbert space of the first and second photon in the pair, respectively. For instance, $|\zeta_1\rangle$ may represent the path state of one photon, while $|\eta_2\rangle$ is the polarization state of the other photon in the pair.

For biphotons with coupled DoFs, on the other hand, the entire biphoton wavepacket in one DoF is coupled or entangled to that in the other DoF. A simple example of such a state, which is also considered in this work, can be expressed as:
\begin{equation}
|\psi_\text{coupled}\rangle=\frac{1}{\sqrt{2}}(|\psi_1^\text{DoF 1}\rangle\otimes|\psi_1^\text{DoF 2}\rangle+|\psi_2^\text{DoF 1}\rangle\otimes|\psi_2^\text{DoF 2}\rangle),
\label{eq3}
\end{equation}
where $|\psi_{i}^{\text{DoF}\,m}\rangle (i=1,2; m=1,2)$ is the entire biphoton wavepacket in DoF $m$, which is also orthogonal to $|\psi_{j}^{\text{DoF}\,m}\rangle (i\neq j)$. This class of biphoton states cannot be factorized into the tensor product in different DoFs (unlike hyper-entangled biphotons). 

\begin{figure}[tpb]
\centering
\includegraphics[width=\linewidth]{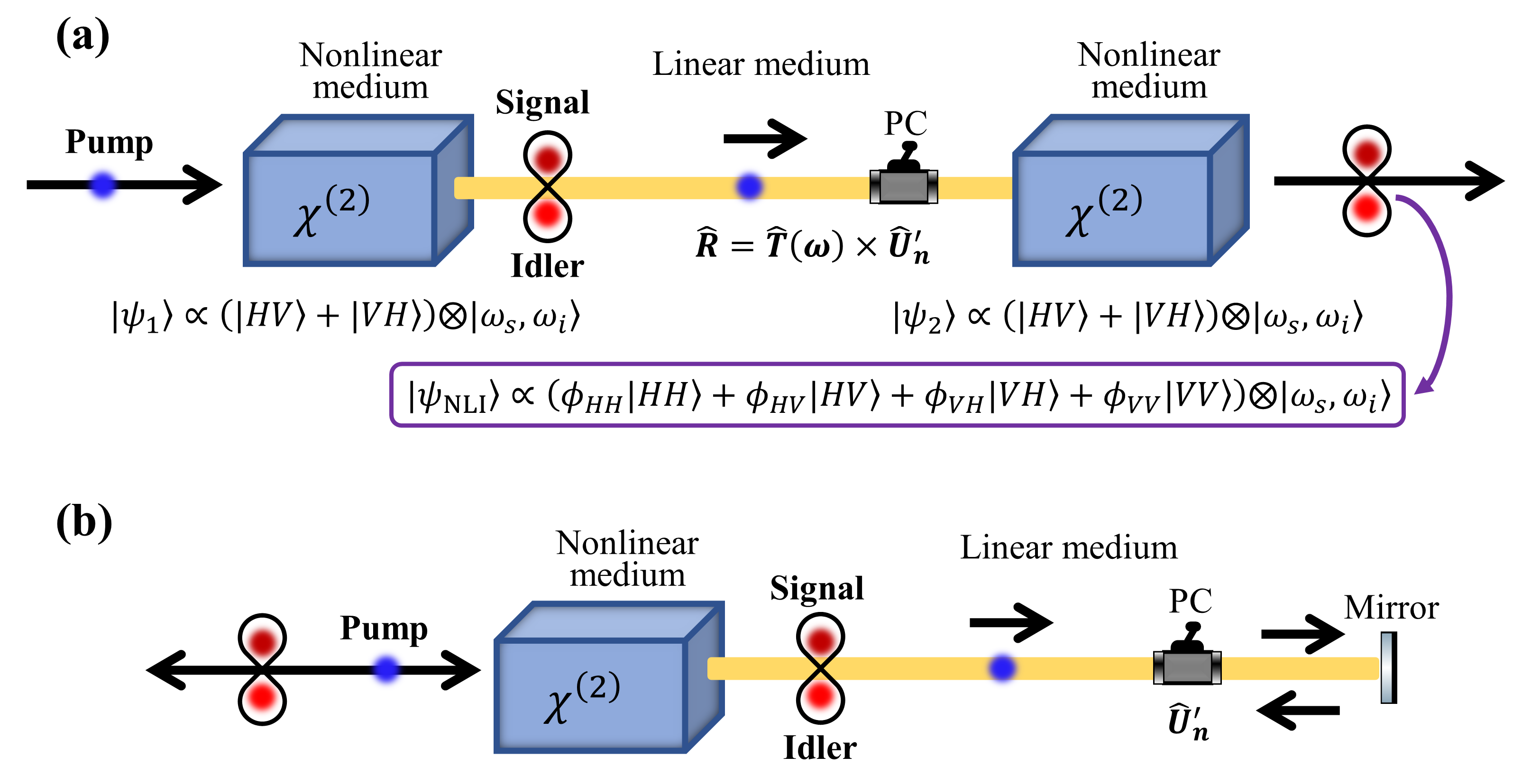}
\caption{(a) General scheme of a common-path nonlinear interferometer (CP-NLI) consisting of two nonlinear media, pumped to generate polarization-entangled biphotons. The wavefunction of the biphotons from the first nonlinear medium $|\psi_1\rangle$ undergoes a transformation $\hat{R}$ and interferes with the biphoton wavefunction from the second nonlinear medium $|\psi_2\rangle$. The transformation $\hat{R}$ determines the output state $|\psi_\text{NLI}\rangle$. (b) A modified reflective CP-NLI. The biphoton amplitudes generated from the forward and backward paths interfere with each other. PC: polarization controller.}
\label{fig1}
\end{figure}

The basic idea of generating biphotons with coupled DoFs is based on interfering biphoton amplitudes generated coherently in two nonlinear media which are separated by a linear dispersive medium, as shown in Fig.\,\ref{fig1}(a). The output biphoton wavefunction, assuming equal amplitude contribution from the two nonlinear processes, can be written as:
\begin{equation}
|\psi_\text{NLI}\rangle=\frac{1}{\sqrt{2}}[e^{i\alpha(\omega_A,\omega_B)}|\psi_1\rangle+|\psi_2\rangle]\otimes|\omega_A,\omega_B\rangle,
\label{eq4}
\end{equation}
where $|\psi_1\rangle$ and $|\psi_2\rangle$ are the two interfering biphoton wavefunctions, generated separately in the two nonlinear media, and $\alpha(\omega_A,\omega_B)$ is a frequency-dependent relative phase due to the linear dispersive medium. Note that the wavefunction in Eq.\,(\ref{eq4}) is not normalized.

If $|\psi_1\rangle$ and $|\psi_2\rangle$ are two identical polarization states, such as being equivalent to $|\Psi^+\rangle=\frac{1}{\sqrt{2}}(|HV\rangle+|VH\rangle)$, the output state will be
\begin{equation}
|\psi_\text{NLI}\rangle=\frac{1}{\sqrt{2}}|\Psi^+\rangle\otimes[e^{i\alpha(\omega_A,\omega_B)}+1]|\omega_A,\omega_B\rangle,
\label{eq5}
\end{equation}
which regardless of $\alpha$ is in the polarization state $|\Psi^+\rangle$ with an amplitude modulation dependent on $\alpha$. The spectral intensity $(\langle\psi_\text{NLI}|\psi_\text{NLI}\rangle)$ varies as $\cos^2(\alpha/2)$, which means that the interference fringe can be observed in the frequency DoF of biphotons. Also, it can be easily shown that the polarization and frequency DoFs are decoupled from each other.

Now if $|\psi_1\rangle$ and $|\psi_2\rangle$ are different and orthogonal to each other, e.g., $|\psi_1\rangle=|\Phi^-\rangle=\frac{1}{\sqrt{2}}(|HH\rangle-|VV\rangle)$, and $|\psi_2\rangle=|\Psi^+\rangle$, then 
\begin{equation}
|\psi_\text{NLI}\rangle=\frac{1}{\sqrt{2}}[e^{i\alpha(\omega_A,\omega_B)}|\Phi^-\rangle+|\Psi^+\rangle]\otimes|\omega_A,\omega_B\rangle.
\label{eq6}
\end{equation}
The spectral intensity $(\langle\psi_\text{NLI}|\psi_\text{NLI}\rangle)$ becomes frequency independent, while the polarization state is dependent on $\alpha$ (and hence on frequency). It shows that the degree of polarization entanglement (expressed as concurrence\,\cite{wootters_entanglement_1998}  $\mathcal{C}^\text{Pol}=|\cos\alpha(\omega_A,\omega_B)|$) depends on the frequencies of biphotons, hence the polarization and frequency DoFs are coupled to each other. As a result of this coupling, we can control the entanglement in one DoF (polarization) by varying the other DoF (frequency). 

Since $|\Phi^-\rangle$ can be obtained from $|\Psi^+\rangle$ by a simple polarization rotation, we can always switch from the biphoton state with decoupled DoFs [Eq.\,(\ref{eq5})] to the one with coupled DoFs [Eq.\,(\ref{eq6})] and vice versa. The role of the polarization controller (PC) in Fig.\,\ref{fig1} is to couple or decouple the two DoFs on demand.

\section{Experimental setup}
To demonstrate our approach experimentally, we use a reflective CP-NLI\,\cite{riazi_dispersion_2020} shown in Fig.\,\ref{fig1}(b). The biphoton amplitudes generated by the forward-propagating and back-reflected pump interfere with each other. The reflective CP-NLI automatically satisfies the requirement of having identical biphoton emission spectra for the two interfering biphoton wavefunctions. Furthermore, since the loss (waveguide loss \& reflection loss) are approximately the same for the pump and the signal/idler, this reflective CP-NLI configuration also automatically ensures that the interfering biphoton wavefunctions have approximately equal amplitudes.

\begin{figure}[htpb]
\centering
\includegraphics[width=\linewidth]{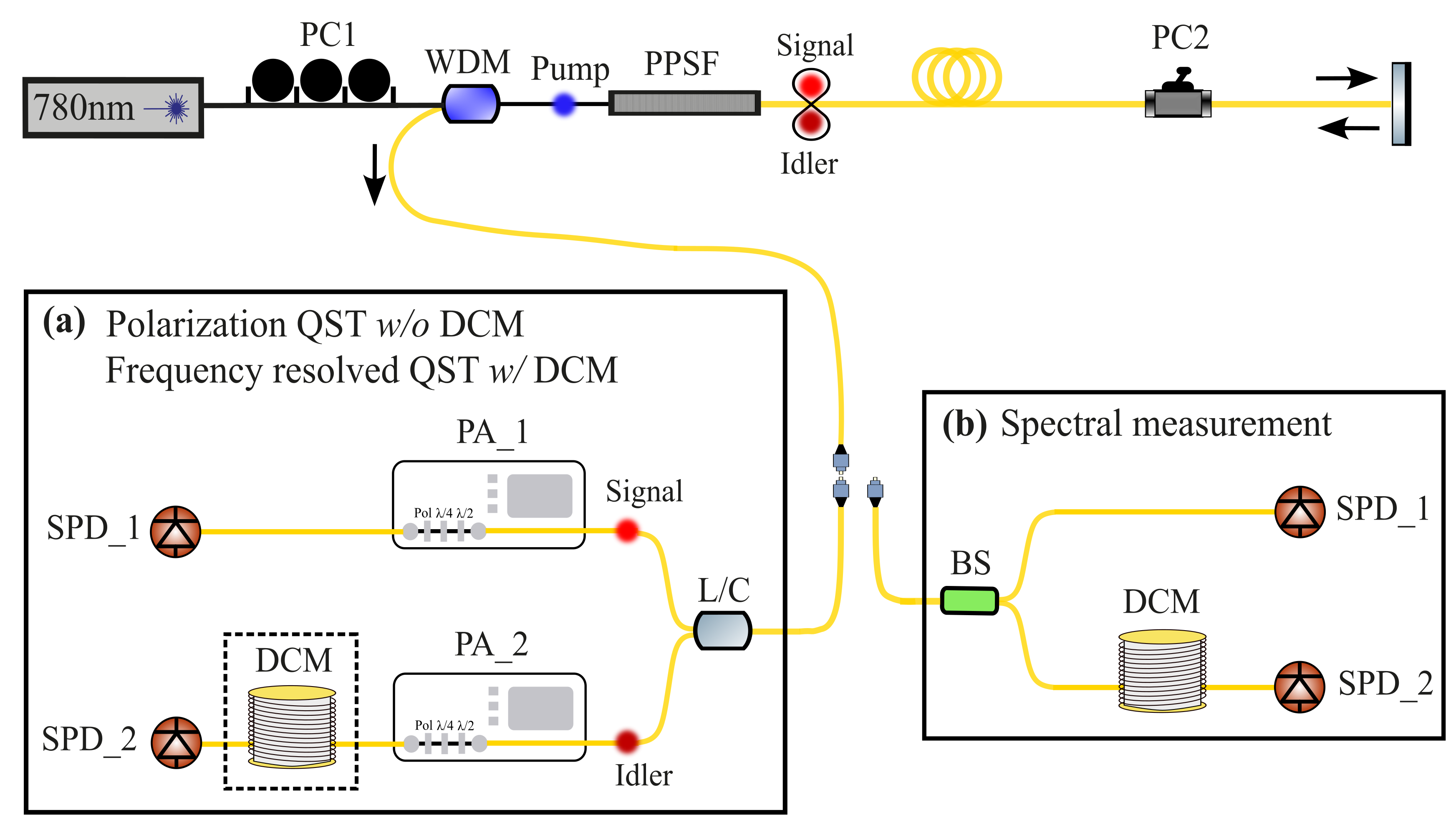}
\caption{Schematic of the experimental setup. The 780 nm pump from a narrow-band laser diode is sent to the PPSF through a wavelength division multiplexer (WDM). The polarization of the pump is controlled by a polarization controller (PC1) for type II phase-matched SPDC. The biphotons generated from the forward and backward paths interfere, producing an output state, which is sent to the measurement apparatus. In inset (a), the L/C band splitter separates signal ($\lambda>$1564 nm) and idler ($\lambda<$1564 nm) into two spatial modes, where we can use polarization analyzers (PA\_1 and PA\_2) to characterize the polarization state through quantum state tomography (QST). This setup is also used to obtain frequency-resolved QST by inserting a dispersion compensating module (DCM). In inset (b), we use this DCM to temporally disperse the biphotons for spectral measurements\,\cite{riazi_biphoton_2019}.}
\label{fig2}
\end{figure}

Our experimental setup is illustrated in Fig.\,\ref{fig2}. The reflective CP-NLI consists of an approximately 20-cm-long periodically-poled silica fiber (PPSF) \cite{Helt_proposal_2009} serving as a nonlinear medium, an approximately 5m-long SMF-28 fiber functioning as a linear dispersive element, and a fiber-based polarization controller used to perform polarization transformation. The reflector is implemented via metal coating of one end of the SMF-28 fiber connector. The nonlinear medium PPSF is used for generating broadband polarization-entangled biphotons ($|\Psi^+\rangle$)\,\cite{zhu_direct_2012,chen_compensation-free_2017,chen_turn-key_2018}. Its broad emission bandwidth and negligible group birefringence significantly simplify the construction of an all-fiber reflective CP-NLI. To produce type-II SPDC, we employ a continuous-wave 780 nm diode laser (Toptica DL-Pro) as the pump. The coherence length of the pump is much greater than the group delay between the pump and signal/idler, allowing us to generate coherent biphotons in maximally polarization-entangled states in both the forward and backward directions. 

To demonstrate the coupling between the frequency and polarization DoFs, we measure the degree of polarization entanglement (more specifically, concurrence), as a function of signal/idler frequencies. To obtain the density matrix for each signal and idler frequency pair, we incorporate a dispersion compensating module (DCM) which maps the spectrum into the delay of the arrival time of signal and idler photons. We perform standard 16-measurement QST and employ the maximum likelihood method to derive the density matrix for each wavelength. The DCM maintains an approximately constant nominal dispersion of around 707 ps/nm. The shortest coincidence window is dictated by the time jitter of the detectors $\sim$256 ps, which gave us a wavelength resolution of approximately 0.36 nm (50 GHz).

\begin{table*}
\caption{\label{table1}Features of biphoton states with decoupled and coupled DoFs.}
\SetTblrInner{rowsep=3pt}
\begin{tblr}{Q[c,3cm]Q[c,4cm]Q[c,9cm]}
\hline
\hline
 & \textbf{Case 1} &  \textbf{Case 2} \\
 \hline
 $\hat{U}_{A(B)}$ & $\begin{pmatrix} 1 & 0 \\ 0 & 1 \end{pmatrix}$ & $ \frac{1}{\sqrt{2}}\begin{pmatrix} 1 & -1 \\ 1 & 1 \end{pmatrix}$ \\
 \hline
Schmidt form & $|\xi_1\rangle_\text{Pol}\otimes|\chi_1\rangle_\text{Freq}$ & $c_1|\xi_1\rangle_\text{Pol}\otimes|\chi_1\rangle_\text{Freq}+c_2|\xi_2\rangle_\text{Pol}\otimes|\chi_2\rangle_\text{Freq}$ \\
\hline
Schmidt rank & 1 & 2 \\
\hline
$\hat{\rho}'(\omega_A,\omega_B)$  & $\frac{1}{2}\begin{pmatrix} 0 & 0 & 0 & 0 \\ 0 & 1 & 1 & 0 \\ 0 & 1 & 1 & 0 \\0 & 0 & 0 & 0  \end{pmatrix}$ & $\frac{1}{4}\begin{pmatrix} 1 & -\rho^*_0(\omega_A,\omega_B) & -\rho^*_0(\omega_A,\omega_B) & -1 \\ -\rho_0(\omega_A,\omega_B) & 1 & 1 & \rho_0(\omega_A,\omega_B) \\ -\rho_0(\omega_A,\omega_B) & 1 & 1 & \rho_0(\omega_A,\omega_B) \\-1 & \rho^*_0(\omega_A,\omega_B) & \rho^*_0(\omega_A,\omega_B) & 1  \end{pmatrix}$ \\
\hline
Concurrence & $\mathcal{C}^\text{Pol}(\omega_A,\omega_B)=1$ & $\mathcal{C}^\text{Pol}(\omega_A,\omega_B)=|\cos2\alpha(\omega_A,\omega_B)|$ \\
\hline
\hline
\end{tblr}
\end{table*}

\section{Results and Discussion}
First, we give a brief derivation of the output quantum states from the CP-NLI for both the decoupled and coupled cases. The general quantum state of biphotons at the output of the CP-NLI in Fig.\,\ref{fig1} is derived in ref.\,\cite{riazi_biphoton_2019}, which considers both the polarization and frequency DoFs, as well as pump coherence. 
After pumped for type-II SPDC, the quantum state of biphotons generated from a nonlinear medium of a length $L$ is:
\begin{equation}
\begin{aligned}
|\psi\rangle= & \int\limits\mathrm{d}\omega_A\mathrm{d}\omega_B\phi(\omega_A,\omega_B) e^{i\Delta k_{VHV}L}[\hat{a}^\dagger_{AH}\hat{a}^\dagger_{BV} \\
& +e^{i\Lambda}\hat{a}^\dagger_{AV}\hat{a}^\dagger_{BH}]|vac\rangle,
\end{aligned}
\label{eq7}
\end{equation}
where $\hat{a}^\dagger_{AS'}(\hat{a}^\dagger_{BS''})$ is the creation operator of the signal (idler) field with $S'(S^{''})$ polarization; $\Delta k_{VVH} = k_{\omega_{PV}}-k_{\omega_{AV}}-k_{\omega_{BH}}-k_{QPM}$ is the phase mismatch between the pump ($V$), signal ($V$) and idler ($H$) in the quasi-phase-matched (QPM) PPSF structure, and $\Delta k_{VHV}$ is similarly defined; $\Lambda=[\Delta k_{VVH}(\omega_A,\omega_B)-\Delta k_{VHV}(\omega_A,\omega_B)]L$ is related to the group birefringence of the PPSF \cite{chen_compensation-free_2017}.
The biphoton amplitude is defined as:
\begin{equation}
\begin{aligned}
\phi(\omega_A,\omega_B) = & \int\limits\mathrm{d}\omega_P\mathcal{A}_0(\omega_P,\omega_A,\omega_B)f(\omega_P)L \\
& \times\sinc\Big(\frac{\Delta kL}{2}\Big)\delta(\omega_P-\omega_A-\omega_B),
\end{aligned}
\label{eq8}
\end{equation}
where $f(\omega_P)$ is the pump linewidth, $\mathcal{A}_0(\omega_P,\omega_A,\omega_B)$ includes the nonlinear susceptibilities and other phase factors. To simplify the analysis, we made the approximation $\Delta k_{VVH}\approx\Delta k_{VHV}$ across the entire phase-matching bandwidth and substituted them with $\Delta k$. The approximation is justified due to the extremely low group birefringence of the PPSF\,\cite{chen_compensation-free_2017} and leads to
 $\Lambda\ll1$. Consequently, the same spectral amplitude is used for the $|HV\rangle_{\omega_A,\omega_B}$ and $|VH\rangle_{\omega_A,\omega_B}$ states in Eq.\,(\ref{eq7}).

The coherent superposition of the two biphoton amplitudes results in the following generic form:
\begin{equation}
\begin{aligned}
|\psi_\text{NLI}\rangle = & \int\limits\mathrm{d}\omega_A \mathrm{d}\omega_B[\phi_{HH}(\omega_A,\omega_B)|HH\rangle \\
& +\phi_{HV}(\omega_A,\omega_B)|HV\rangle \\
& +\phi_{VH}(\omega_A,\omega_B)|VH\rangle \\
& +\phi_{VV}(\omega_A,\omega_B)|VV\rangle]\otimes|\omega_A,\omega_B\rangle,
\end{aligned}
\label{eq9}
\end{equation}
where $\phi_{S'S''}(\omega_A,\omega_B)$ is the biphoton amplitude of each polarization state. Note that $\phi_{S'S''}(\omega_A,\omega_B)$ includes the effects of both dispersions of the linear medium and polarization transformation of PC2. Their collective effect can be modeled by two consecutive transformations: 
\begin{equation}
\hat{R}_n=\hat{T}(\omega_n)\times\hat{U}_n(\theta_n,\phi_{1n},\phi_{2n}),
\label{eq10}
\end{equation}
where $\hat{T}(\omega_n)=e^{2ik_n(\omega_n)L_0}\hat{\mathbb{I}}$ is the phase accumulation due to dispersion of the linear medium. The generic unitary transformation for a PC is
\begin{equation}
\hat{U}_n(\theta_n,\phi_{1n},\phi_{2n})=\begin{pmatrix} e^{i\phi_{1n}}\cos\theta_n & -e^{i\phi_{2n}}\sin\theta_n \\ e^{-i\phi_{2n}}\sin\theta_n & e^{-i\phi_{1n}}\cos\theta_n \end{pmatrix},
\label{eq11}
\end{equation}
The subscript $n$ can be $P$, $A$, and $B$, for pump, signal, and idler, respectively. Note that $\hat{U}_n$ is assumed to be weakly wavelength-dependent, such that $\hat{U}_A=\hat{U}_B\neq\hat{U}_P$.

The polarization transformation plays an essential role as it is responsible for coupling the frequency and polarization DoFs of biphotons. It is assumed that PC2 has broad spectral coverage over the signal and idler bandwidth, therefore the angles $\theta$, $\phi_1$, and $\phi_2$ are the same for both photons. The values of $\phi_1$ and $\phi_2$ are determined by the birefringence introduced by the PC2. For simplicity, we set $\phi_1 = \phi_2 = 0$, but our findings can be easily generalized to cases with non-zero $\phi_1$ and $\phi_2$\,\cite{riazi_biphoton_2019}. To demonstrate this effect, we consider two special cases shown in Table\,\ref{table1}. For \textbf{Case 1}, there is no polarisation rotation ($\theta=0$) and the two DoFs of biphotons become \textit{decoupled}. For \textbf{Case 2}, the polarisation is rotated by $\theta=\pi/4$ and the two DoFs of biphotons become \textit{coupled}.

To illustrate the coupling between the two DoFs, we first consider Schmidt decomposition of the biphoton states into frequency and polarization, i.e. $|\psi_\text{NLI}\rangle=\sum_k|\xi_k\rangle_\text{Pol}|\chi_k\rangle_\text{Freq}$. Note that the definition of subsystems we use here differs from the one normally used in the literature, where the two subsystems consist of signal and idler photons. However, since our focus is on the coupling between two DoFs of biphotons, we consider the complete biphoton state in both polarization $|\xi_k\rangle_\text{Pol}$ and frequency $|\chi_k\rangle_\text{Freq}$ DoFs as our subsystems for decomposition. With this definition of subsystems, biphoton states with Schmidt rank greater than 1 indicate that the two subsystems are entangled, implying a coupling between the two DoFs. Consequently, apart from the entanglement between the two photons in the polarization DoF, the two DoFs themselves are also coupled.

\begin{figure*}[htpb]
\centering
\includegraphics[width=0.9\linewidth]{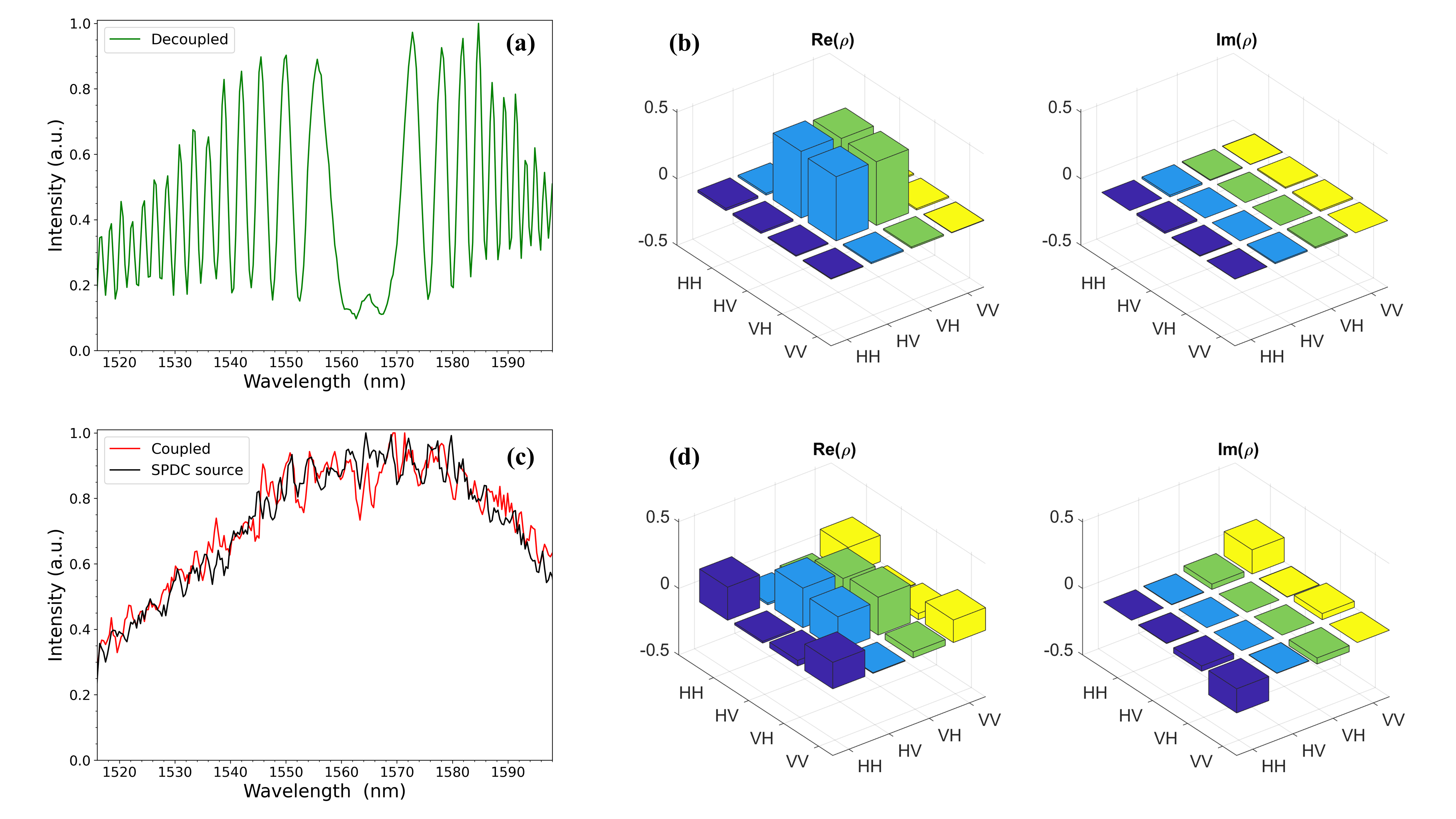}
\caption{Experimentally measured biphoton properties in polarization and frequency DoFs. In \textbf{Case 1} (no polarization rotation) the two DoFs are decoupled: (a) tracing over the polarization DoF has no effect on the interference fringes observed in the spectrum; (b) tracing over frequency DoF also has no effect on the degree of polarization entanglement [$\mathcal{C}^\text{Pol}=0.95(2)$]. In \textbf{Case 2} (a polarization rotation of $\pi/4$) the two DoFs are coupled: (c) tracing over the polarization DoF completely destroys the interference in the spectrum. The measured spectrum of the biphoton state at the output of the NLI is identical to that from the SPDC source. (d) Tracing over frequency DoF leads to the depletion of polarization entanglement [$\mathcal{C}^\text{Pol}=0.10(4)$]. The cut-off wavelengths of the DCM constrain the range of the spectrum.}
\label{fig3}
\end{figure*}

\subsection{\textbf{Case 1}: \textit{biphotons with decoupled DoFs}}
When there is no polarization rotation applied to the biphoton amplitude, the quantum state at the output of the CP-NLI can be derived as:
\begin{equation}
\begin{aligned}
|\psi_\text{NLI}\rangle_\text{case1} = & \int\limits\mathrm{d}\omega_A \mathrm{d}\omega_B\frac{1}{2}[|HV\rangle+|VH\rangle]\otimes \\
& \phi(\omega_A,\omega_B)[1+e^{2i\alpha(\omega_A,\omega_B)}]|\omega_A,\omega_B\rangle,
\end{aligned}
\label{eq12}
\end{equation}
where $\alpha(\omega_A,\omega_B)=\Delta k^{(0)}(\omega_A,\omega_B)L_0+\Delta k(\omega_A,\omega_B)L$, the phase mismatch of the pump, signal, and idler fields in the linear and nonlinear medium, respectively; $\phi(\omega_A,\omega_B)$ is the biphoton wavefunction. The rank-1 biphoton state can be written in Schmidt form 
\begin{equation}
|\psi_\text{NLI}\rangle_\text{case1} = |\xi_1\rangle_\text{Pol}\otimes|\chi_1\rangle_\text{Freq},
\label{eq13}
\end{equation}
where $|\xi_1\rangle_\text{Pol}=|\Psi^+\rangle$ and $|\chi_1\rangle_\text{Freq}=\int\limits\mathrm{d}\omega_A \mathrm{d}\omega_B\phi(\omega_A,\omega_B)[1+e^{2i\alpha(\omega_A,\omega_B)}]|\omega_A,\omega_B\rangle$, implies that the polarization and frequency DoFs are completely decoupled from each other. 

In order to gain further insight into the biphoton state, we analyze its spectral intensity by tracing over the polarization DoF. The spectral intensity of biphoton state $|\psi_\text{NLI}\rangle_\text{case1}$ is proportional to $|\phi(\omega_A,\omega_B)|^2\cos^2\alpha(\omega_A,\omega_B)$\,\cite{riazi_biphoton_2019}, which results in the appearance of spectral fringes. These fringes are generated by the interference between forward- and backward-propagating biphoton amplitudes and are associated with the spectral phase $\alpha(\omega_A,\omega_B)$. By employing the spectral measurement apparatus illustrated in Fig.\,\ref{fig2}(b) for coincidence detection in a polarization-independent manner, we experimentally observe this phenomenon, as shown in Fig.\,\ref{fig3}. Figure\,\ref{fig3}(a) clearly shows that tracing over the polarization DoF does not affect the frequency DoF of biphotons, and the two biphoton amplitudes continue to spectrally interfere with each other as expected.

Our focus now shifts to examining the polarization DoF of biphotons. Trace over the frequency DoF and generate the reduced density matrix of the biphoton state:
\begin{equation}
\begin{aligned}
\hat{\rho}^\text{red-pol}_\text{case1} & = Tr_\text{Freq}(|\psi_\text{NLI}\rangle\langle\psi_\text{NLI}|) \\
& = \int\limits\mathrm{d}\omega_A \mathrm{d}\omega_B f(\omega_A,\omega_B)\times\hat{\rho}'_\text{case1}(\omega_A,\omega_B),
\end{aligned}
\label{eq14}
\end{equation}
where $f(\omega_A,\omega_B)=|\tilde\phi(\omega_A,\omega_B)|^2$ (with $\tilde\phi(\omega_A,\omega_B)=\phi(\omega_A,\omega_B)[1+e^{2i\alpha(\omega_A,\omega_B)}]$) is a slow-varying frequency-dependent factor that is not of importance here. $\hat{\rho}'_\text{case1}(\omega_A,\omega_B)$ is the density matrix corresponding to each signal and idler frequency-conjugate pair. For the biphotons with decoupled DoFs, $\hat{\rho}'_\text{case1}(\omega_A,\omega_B)$ represent a maximally polarization entangled state (concurrence $\mathcal{C}(\omega_A,\omega_B)=1$) and is independent of signal and idler frequencies. Upon evaluating the integral in Eq.\,(\ref{eq14}), the reduced polarization density matrix $\hat{\rho}^\text{red-pol}_\text{case1}$ is obtained, as presented in Table \ref{table1}. The coherence elements in $\hat{\rho}^\text{red-pol}_\text{case1}$ indicate that quantum interference has not been destroyed by tracing over the frequency DoF. As a consequence, the entire biphoton wavepacket remains in a maximally-entangled polarization state, which has been confirmed experimentally using the QST measurement apparatus illustrated in Fig.\,\ref{fig2}(a). The level of polarization entanglement, as observed in the experimentally acquired polarization density matrix shown in Fig.\,\ref{fig3}(b), is of high degree [$\mathcal{C}^\text{Pol} = 0.95(2)$]. It is significant to note that the coherence in the polarization DoF is preserved even after integrating over the signal and idler frequencies, due to the decoupling of the two DoFs.

\begin{figure*}[!t]
\centering
\includegraphics[width=0.85\linewidth]{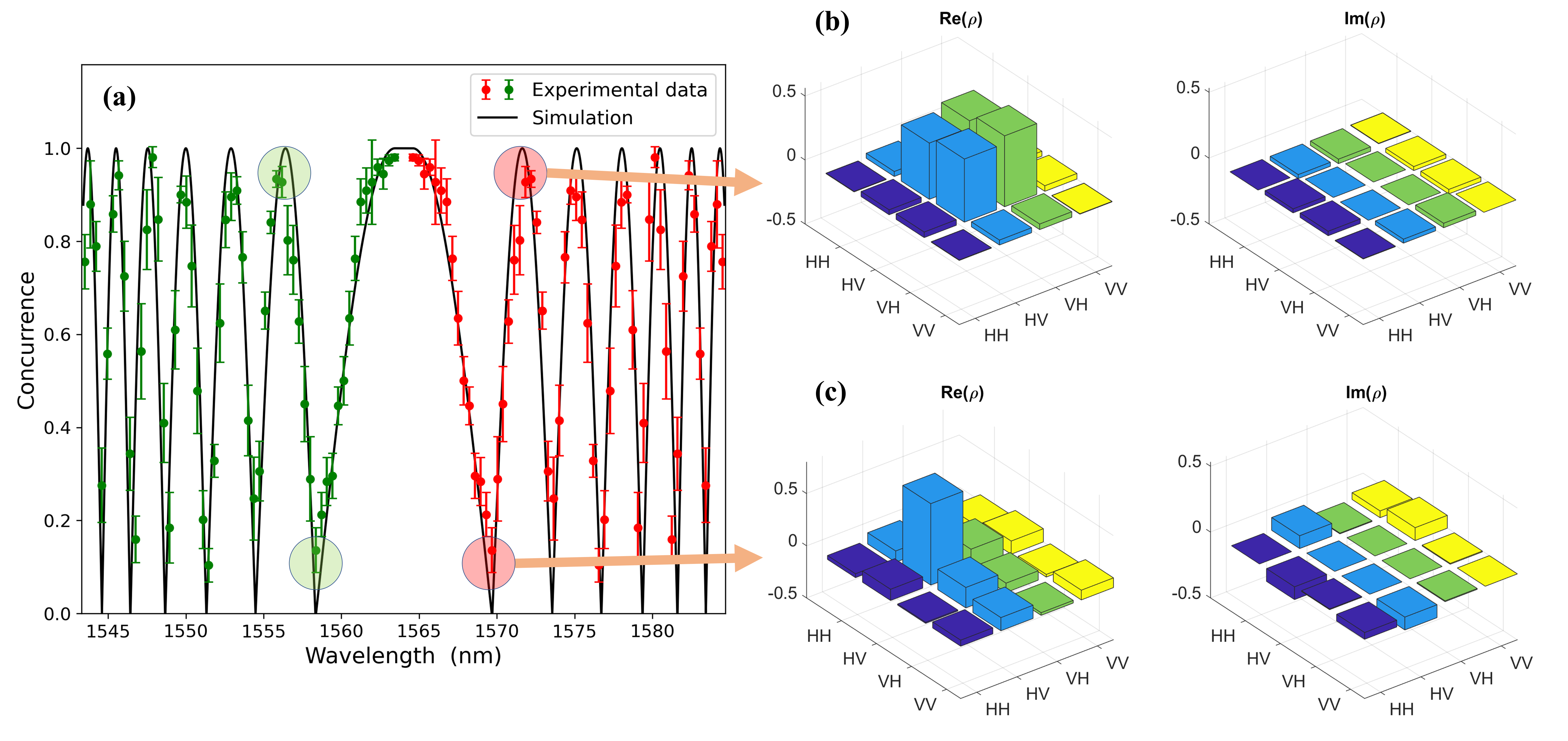}
\caption{Frequency-resolved quantum-state tomography. (a) The concurrence is dependent on the signal/idler photon frequencies (wavelengths) when the polarization and frequency DoFs are coupled. The polarization state density matrix reconstructed from the polarization quantum-state tomography with (b) high and (c) low concurrence.}
\label{fig4}
\end{figure*}

\subsection{\textbf{Case 2}: \textit{biphotons with coupled DoFs}}
When a polarization rotation of $\pi/4$ is applied to the biphoton amplitude,
\begin{equation}
\begin{aligned}
|\psi_\text{NLI}\rangle_\text{case2} = & \frac{1}{2}\int\limits\mathrm{d}\omega_A \mathrm{d}\omega_B[-|HH\rangle +|VV\rangle \\
& +e^{2i\alpha(\omega_A,\omega_B)}(|HV\rangle + |VH\rangle)] \\
& \otimes\phi(\omega_A,\omega_B)|\omega_A,\omega_B\rangle.
\end{aligned}
\label{eq15}
\end{equation}
In contrast to Eq.\,(\ref{eq12}), the frequency and polarization DoFs cannot be separated due to their coupling. In fact, the Schmidt decomposition of the state in Eq.\,(\ref{eq15}) yields:
\begin{equation}
|\psi_\text{NLI}\rangle_\text{case2} = \sum_{k=1}^2 c_k|\xi_k\rangle_\text{Pol}\otimes|\chi_k\rangle_\text{Freq},
\label{eq16}
\end{equation}
where $c_1=c_2=\frac{1}{\sqrt{2}}$, $|\xi_1\rangle_\text{Pol}=-|\Phi^-\rangle$, $|\xi_2\rangle_\text{Pol}=|\Psi^+\rangle$, 
$|\chi_1\rangle_\text{Freq}=\int\limits\mathrm{d}\omega_A\mathrm{d}\omega_B\phi(\omega_A,\omega_B)|\omega_A,\omega_B\rangle$ and $|\chi_2\rangle_\text{Freq}=\int\limits\mathrm{d}\omega_A\mathrm{d}\omega_Be^{2i\alpha(\omega_A,\omega_B)}\phi(\omega_A,\omega_B)|\omega_A,\omega_B\rangle$. Since the Schmidt rank for $|\psi_\text{NLI}\rangle_\text{case2}$ is 2 and $c_1=c_2$, the frequency and polarization DoFs are maximally coupled.

Following a similar approach, we trace over the polarization DoF and analyze the spectrum of the biphotons. The absence of interference between the biphoton amplitudes from the forward and backward paths leads to the disappearance of spectral fringes. Using the spectral measurement apparatus depicted in Fig.\,\ref{fig2}(b), there is no modulation of spectral intensity for \textbf{Case 2}, as can be seen in Fig.\,\ref{fig3}(c).

Likewise, we trace over the frequency DoF and form the reduced density matrix: 
\begin{equation}
\begin{aligned}
\hat{\rho}^\text{red-pol}_\text{case2} & = Tr_\text{Freq}(|\psi_\text{NLI}\rangle\langle\psi_\text{NLI}|) \\
& = \int\limits\mathrm{d}\omega_A \mathrm{d}\omega_B g(\omega_A,\omega_B)\times\hat{\rho}'_\text{case2}(\omega_A,\omega_B),
\end{aligned}
\label{eq17}
\end{equation}
where $g(\omega_A,\omega_B) = |\phi(\omega_A,\omega_B)|^2$ is also a slow-varying frequency-dependent factor. $\hat{\rho}'_\text{case2}(\omega_A,\omega_B)$ is the density matrix presented in Table \ref{table1}, where $\rho_0(\omega_A,\omega_B)=e^{2i\alpha(\omega_A,\omega_B)}$ is a function of the dispersion of the linear material. Calculating the integral in Eq.\,(\ref{eq17}), we obtain a reduced polarization density matrix:
\begin{equation}
\hat{\rho}^\text{red-pol}_\text{case2} = \frac{1}{4}\begin{pmatrix} 1 & 0 & 0 & -1 \\ 0 & 1 & 1 & 0 \\ 0 & 1 & 1 & 0 \\-1 & 0 & 0 & 1  \end{pmatrix},
\label{eq18}
\end{equation}
which is a mixed state: $\hat{\rho}^\text{red-pol}_\text{case2}=\frac{1}{2}(|\Psi^+\rangle\langle\Psi^+|+|\Phi^-\rangle\langle\Phi^-|)$. A low concurrence of $\mathcal{C}^\text{Pol} = 0.10(4)$ was measured, as shown in Fig.\,\ref{fig3}(d). It should be noted that the experimental concurrence is not reduced to zero, as the manual PC has limited tuning accuracy and the biphoton amplitudes from the two paths are not equal. The results depicted in Fig.\,\ref{fig3}(c,d) support the coupling between the two DoFs of biphotons, as tracing over one DoF affects the other DoF by destroying the quantum interference.
One can clearly see that $|\psi_\text{NLI}\rangle_\text{case1}$ is always a maximally polarization entangled, regardless of $\omega_A$ and $\omega_B$. However, the polarization state of $|\psi_\text{NLI}\rangle_\text{case2}$ depends on signal and idler frequencies $\omega_A$ and $\omega_B$.

The unique signature of the biphoton state in Eq.\,(\ref{eq17}) is that the density matrix $\hat{\rho}'_\text{case2}(\omega_A,\omega_B)$ is frequency dependent for each signal and idler frequency-conjugate pair. This is a result of the coupling between the polarization and frequency DoFs of the entire biphoton wavepacket. To observe this property, we spectrally resolve the concurrence by performing QST\,\cite{james_measurement_2001} on many frequency-conjugate pairs (over a 60 nm bandwidth). Figure\,\ref{fig4} show that the concurrence of the biphotons is frequency-dependent, ranging from 0.12(5) to 0.98(2), with the variation given by $\mathcal{C}^\text{Pol}=|\cos2\alpha(\omega_A,\omega_B)|$. In other words, the degree of polarization entanglement of biphotons can be controlled by tuning their frequencies. 
Although the observed fringes in Fig.\,\ref{fig4}(a) are related to those in Fig.\,\ref{fig3}(a), it is important to note that the former exhibits fringes in concurrence while the latter displays fringes in spectral brightness. This means that the polarization rotation of $\pi/4$ converts the modulation of brightness to the modulation of concurrence. Moreover, the modulation pattern observed in Fig.\,\ref{fig4} depends on the dispersion of the linear medium $\alpha(\omega_A,\omega_B)$ and can be tailored accordingly.

The reflective CP-NLI enables us to tune concurrence from 0 to 1 by selecting appropriate frequency pairs. Furthermore, as concurrence is invariant under local unitary operations, local polarization transformations can be combined with our system to obtain any arbitrary pure biphoton polarization state, ranging from a completely separable state to a maximally entangled state. It is worth noting that a broad biphoton bandwidth is not always necessary for polarization entanglement tuning, although it does make frequency-dependent concurrence obtainable with only a modest amount of dispersion. Equation\,(\ref{eq6}) suggests that polarization entanglement can be controlled by the interference between two maximally-entangled Bell states ($|\Phi^-\rangle$ and $|\Psi^+\rangle$) with varying relative phase $\alpha(\omega_A,\omega_B)$. For the case of narrowband biphotons, where it is not practical to use the frequency-dependent phase for entanglement tuning, phase modulators can still be used to arbitrarily vary $\alpha$ and control the entanglement or obtain any arbitrary biphoton polarization state in a CP-NLI.

Our theoretical analysis and experimental results demonstrate the coupling and decoupling phenomena by observing the frequency (or polarization) while tracing over the other DoF. When the two DoFs are coupled, tracing over one of them results in the destruction of quantum interference in the other DoF. The loss of spectral fringes is a clear indication of the destruction of quantum interference in the frequency DoF, while a biphoton polarization state with low concurrence is an indication of the destruction of quantum interference in the polarization DoF.

\section{Conclusion}
We introduce a novel approach for coupling and decoupling two DoFs (polarization and frequency) of biphotons through a reflective CP-NLI. In contrast to previous methods\,\cite{guo_frequency-bin_2015,shu_narrowband_2015}, our approach eliminates the need for phase stabilization or active element for DoF coupling. 
We further experimentally demonstrated the on-demand coupling/decoupling of the two DoFs through a straightforward polarization transformation within the CP-NLI. Once the two DoFs of biphotons were decoupled, the interference fringes were initially observed in their spectrum. As a consequence of coupling the two DoFs, the interference fringes disappeared and were transferred to the polarization domain and resulting in frequency-dependent polarization entanglement. This feature could be utilized to adjust the polarization entanglement by selecting different frequency-conjugate pairs of biphotons. 
Moreover, we anticipate that the CP-NLI approach can be extended to other DoFs, such as time bins, spatial modes, and orbital angular momentum. 
The class of biphotons with coupled DoFs introduced here, combined with the technique mentioned in Ref.\,\cite{chen_recovering_2020}, can be further exploited for the generation of two-photon cluster states\,\cite{vallone_realization_2007,chen_experimental_2007}, which has unique applications in quantum information processing\,\cite{chen_experimental_2007,reimer_high-dimensional_2019}.
Overall, our approach offers a convenient and flexible means of controlling DoFs and could find applications in various fields.

\begin{acknowledgments}
We thank John Sipe for his insightful discussions. This research is supported by NSERC Discovery Grant (RGPIN-2019-07019), NSERC Alliance Grant (ALLRP 569583 EU-Canada joint project HyperSpace), and DEVCOM Army Research Laboratory Grant W911NF-20-2-0242. The equipment used in this research is funded by Canada Foundation for Innovation (CFI). 
\end{acknowledgments}

\bibliography{refs}

\begin{thebibliography}{40}%
\makeatletter
\providecommand \@ifxundefined [1]{%
 \@ifx{#1\undefined}
}%
\providecommand \@ifnum [1]{%
 \ifnum #1\expandafter \@firstoftwo
 \else \expandafter \@secondoftwo
 \fi
}%
\providecommand \@ifx [1]{%
 \ifx #1\expandafter \@firstoftwo
 \else \expandafter \@secondoftwo
 \fi
}%
\providecommand \natexlab [1]{#1}%
\providecommand \enquote  [1]{``#1''}%
\providecommand \bibnamefont  [1]{#1}%
\providecommand \bibfnamefont [1]{#1}%
\providecommand \citenamefont [1]{#1}%
\providecommand \href@noop [0]{\@secondoftwo}%
\providecommand \href [0]{\begingroup \@sanitize@url \@href}%
\providecommand \@href[1]{\@@startlink{#1}\@@href}%
\providecommand \@@href[1]{\endgroup#1\@@endlink}%
\providecommand \@sanitize@url [0]{\catcode `\\12\catcode `\$12\catcode
  `\&12\catcode `\#12\catcode `\^12\catcode `\_12\catcode `\%12\relax}%
\providecommand \@@startlink[1]{}%
\providecommand \@@endlink[0]{}%
\providecommand \url  [0]{\begingroup\@sanitize@url \@url }%
\providecommand \@url [1]{\endgroup\@href {#1}{\urlprefix }}%
\providecommand \urlprefix  [0]{URL }%
\providecommand \Eprint [0]{\href }%
\providecommand \doibase [0]{http://dx.doi.org/}%
\providecommand \selectlanguage [0]{\@gobble}%
\providecommand \bibinfo  [0]{\@secondoftwo}%
\providecommand \bibfield  [0]{\@secondoftwo}%
\providecommand \translation [1]{[#1]}%
\providecommand \BibitemOpen [0]{}%
\providecommand \bibitemStop [0]{}%
\providecommand \bibitemNoStop [0]{.\EOS\space}%
\providecommand \EOS [0]{\spacefactor3000\relax}%
\providecommand \BibitemShut  [1]{\csname bibitem#1\endcsname}%
\let\auto@bib@innerbib\@empty
\bibitem [{\citenamefont {Jennewein}\ \emph {et~al.}(2000)\citenamefont
  {Jennewein}, \citenamefont {Simon}, \citenamefont {Weihs}, \citenamefont
  {Weinfurter},\ and\ \citenamefont {Zeilinger}}]{jennewein_quantum_2000}%
  \BibitemOpen
  \bibfield  {author} {\bibinfo {author} {\bibfnamefont {Thomas}\ \bibnamefont
  {Jennewein}}, \bibinfo {author} {\bibfnamefont {Christoph}\ \bibnamefont
  {Simon}}, \bibinfo {author} {\bibfnamefont {Gregor}\ \bibnamefont {Weihs}},
  \bibinfo {author} {\bibfnamefont {Harald}\ \bibnamefont {Weinfurter}}, \ and\
  \bibinfo {author} {\bibfnamefont {Anton}\ \bibnamefont {Zeilinger}},\
  }\bibfield  {title} {\enquote {\bibinfo {title} {Quantum {Cryptography} with
  {Entangled} {Photons}},}\ }\href {\doibase 10.1103/PhysRevLett.84.4729}
  {\bibfield  {journal} {\bibinfo  {journal} {Phys. Rev. Lett.}\ }\textbf
  {\bibinfo {volume} {84}},\ \bibinfo {pages} {4729--4732} (\bibinfo {year}
  {2000})}\BibitemShut {NoStop}%
\bibitem [{\citenamefont {Kues}\ \emph {et~al.}(2017)\citenamefont {Kues},
  \citenamefont {Reimer}, \citenamefont {Roztocki}, \citenamefont {Cortés},
  \citenamefont {Sciara}, \citenamefont {Wetzel}, \citenamefont {Zhang},
  \citenamefont {Cino}, \citenamefont {Chu}, \citenamefont {Little},
  \citenamefont {Moss}, \citenamefont {Caspani}, \citenamefont {Azaña},\ and\
  \citenamefont {Morandotti}}]{kues_-chip_2017}%
  \BibitemOpen
  \bibfield  {author} {\bibinfo {author} {\bibfnamefont {Michael}\ \bibnamefont
  {Kues}}, \bibinfo {author} {\bibfnamefont {Christian}\ \bibnamefont
  {Reimer}}, \bibinfo {author} {\bibfnamefont {Piotr}\ \bibnamefont
  {Roztocki}}, \bibinfo {author} {\bibfnamefont {Luis~Romero}\ \bibnamefont
  {Cortés}}, \bibinfo {author} {\bibfnamefont {Stefania}\ \bibnamefont
  {Sciara}}, \bibinfo {author} {\bibfnamefont {Benjamin}\ \bibnamefont
  {Wetzel}}, \bibinfo {author} {\bibfnamefont {Yanbing}\ \bibnamefont {Zhang}},
  \bibinfo {author} {\bibfnamefont {Alfonso}\ \bibnamefont {Cino}}, \bibinfo
  {author} {\bibfnamefont {Sai~T.}\ \bibnamefont {Chu}}, \bibinfo {author}
  {\bibfnamefont {Brent~E.}\ \bibnamefont {Little}}, \bibinfo {author}
  {\bibfnamefont {David~J.}\ \bibnamefont {Moss}}, \bibinfo {author}
  {\bibfnamefont {Lucia}\ \bibnamefont {Caspani}}, \bibinfo {author}
  {\bibfnamefont {José}\ \bibnamefont {Azaña}}, \ and\ \bibinfo {author}
  {\bibfnamefont {Roberto}\ \bibnamefont {Morandotti}},\ }\bibfield  {title}
  {\enquote {\bibinfo {title} {On-chip generation of high-dimensional entangled
  quantum states and their coherent control},}\ }\href {\doibase
  10.1038/nature22986} {\bibfield  {journal} {\bibinfo  {journal} {Nature}\
  }\textbf {\bibinfo {volume} {546}},\ \bibinfo {pages} {622--626} (\bibinfo
  {year} {2017})}\BibitemShut {NoStop}%
\bibitem [{\citenamefont {Riazi}\ \emph {et~al.}(2019)\citenamefont {Riazi},
  \citenamefont {Chen}, \citenamefont {Zhu}, \citenamefont {Gladyshev},
  \citenamefont {Kazansky}, \citenamefont {Sipe},\ and\ \citenamefont
  {Qian}}]{riazi_biphoton_2019}%
  \BibitemOpen
  \bibfield  {author} {\bibinfo {author} {\bibfnamefont {Arash}\ \bibnamefont
  {Riazi}}, \bibinfo {author} {\bibfnamefont {Changjia}\ \bibnamefont {Chen}},
  \bibinfo {author} {\bibfnamefont {Eric~Y.}\ \bibnamefont {Zhu}}, \bibinfo
  {author} {\bibfnamefont {Alexey~V.}\ \bibnamefont {Gladyshev}}, \bibinfo
  {author} {\bibfnamefont {Peter~G.}\ \bibnamefont {Kazansky}}, \bibinfo
  {author} {\bibfnamefont {J.~E.}\ \bibnamefont {Sipe}}, \ and\ \bibinfo
  {author} {\bibfnamefont {Li}~\bibnamefont {Qian}},\ }\bibfield  {title}
  {\enquote {\bibinfo {title} {Biphoton shaping with cascaded entangled-photon
  sources},}\ }\href {\doibase 10.1038/s41534-019-0188-1} {\bibfield  {journal}
  {\bibinfo  {journal} {npj Quantum Inf.}\ }\textbf {\bibinfo {volume} {5}},\
  \bibinfo {pages} {1--10} (\bibinfo {year} {2019})}\BibitemShut {NoStop}%
\bibitem [{\citenamefont {Williams}\ \emph {et~al.}(2019)\citenamefont
  {Williams}, \citenamefont {Lukens}, \citenamefont {Peters}, \citenamefont
  {Qi},\ and\ \citenamefont {Grice}}]{williams_quantum_2019}%
  \BibitemOpen
  \bibfield  {author} {\bibinfo {author} {\bibfnamefont {Brian~P.}\
  \bibnamefont {Williams}}, \bibinfo {author} {\bibfnamefont {Joseph~M.}\
  \bibnamefont {Lukens}}, \bibinfo {author} {\bibfnamefont {Nicholas~A.}\
  \bibnamefont {Peters}}, \bibinfo {author} {\bibfnamefont {Bing}\ \bibnamefont
  {Qi}}, \ and\ \bibinfo {author} {\bibfnamefont {Warren~P.}\ \bibnamefont
  {Grice}},\ }\bibfield  {title} {\enquote {\bibinfo {title} {Quantum secret
  sharing with polarization-entangled photon pairs},}\ }\href {\doibase
  10.1103/PhysRevA.99.062311} {\bibfield  {journal} {\bibinfo  {journal} {Phys.
  Rev. A}\ }\textbf {\bibinfo {volume} {99}},\ \bibinfo {pages} {062311}
  (\bibinfo {year} {2019})}\BibitemShut {NoStop}%
\bibitem [{\citenamefont {Vallone}\ \emph {et~al.}(2007)\citenamefont
  {Vallone}, \citenamefont {Pomarico}, \citenamefont {Mataloni}, \citenamefont
  {De~Martini},\ and\ \citenamefont {Berardi}}]{vallone_realization_2007}%
  \BibitemOpen
  \bibfield  {author} {\bibinfo {author} {\bibfnamefont {Giuseppe}\
  \bibnamefont {Vallone}}, \bibinfo {author} {\bibfnamefont {Enrico}\
  \bibnamefont {Pomarico}}, \bibinfo {author} {\bibfnamefont {Paolo}\
  \bibnamefont {Mataloni}}, \bibinfo {author} {\bibfnamefont {Francesco}\
  \bibnamefont {De~Martini}}, \ and\ \bibinfo {author} {\bibfnamefont
  {Vincenzo}\ \bibnamefont {Berardi}},\ }\bibfield  {title} {\enquote {\bibinfo
  {title} {Realization and {Characterization} of a {Two}-{Photon}
  {Four}-{Qubit} {Linear} {Cluster} {State}},}\ }\href {\doibase
  10.1103/PhysRevLett.98.180502} {\bibfield  {journal} {\bibinfo  {journal}
  {Phys. Rev. Lett.}\ }\textbf {\bibinfo {volume} {98}},\ \bibinfo {pages}
  {180502} (\bibinfo {year} {2007})}\BibitemShut {NoStop}%
\bibitem [{\citenamefont {Chen}\ \emph {et~al.}(2007)\citenamefont {Chen},
  \citenamefont {Li}, \citenamefont {Zhang}, \citenamefont {Chen},
  \citenamefont {Goebel}, \citenamefont {Chen}, \citenamefont {Mair},\ and\
  \citenamefont {Pan}}]{chen_experimental_2007}%
  \BibitemOpen
  \bibfield  {author} {\bibinfo {author} {\bibfnamefont {Kai}\ \bibnamefont
  {Chen}}, \bibinfo {author} {\bibfnamefont {Che-Ming}\ \bibnamefont {Li}},
  \bibinfo {author} {\bibfnamefont {Qiang}\ \bibnamefont {Zhang}}, \bibinfo
  {author} {\bibfnamefont {Yu-Ao}\ \bibnamefont {Chen}}, \bibinfo {author}
  {\bibfnamefont {Alexander}\ \bibnamefont {Goebel}}, \bibinfo {author}
  {\bibfnamefont {Shuai}\ \bibnamefont {Chen}}, \bibinfo {author}
  {\bibfnamefont {Alois}\ \bibnamefont {Mair}}, \ and\ \bibinfo {author}
  {\bibfnamefont {Jian-Wei}\ \bibnamefont {Pan}},\ }\bibfield  {title}
  {\enquote {\bibinfo {title} {Experimental {Realization} of {One}-{Way}
  {Quantum} {Computing} with {Two}-{Photon} {Four}-{Qubit} {Cluster}
  {States}},}\ }\href {\doibase 10.1103/PhysRevLett.99.120503} {\bibfield
  {journal} {\bibinfo  {journal} {Phys. Rev. Lett.}\ }\textbf {\bibinfo
  {volume} {99}},\ \bibinfo {pages} {120503} (\bibinfo {year}
  {2007})}\BibitemShut {NoStop}%
\bibitem [{\citenamefont {Shalm}\ \emph {et~al.}(2015)\citenamefont {Shalm},
  \citenamefont {Meyer-Scott}, \citenamefont {Christensen}, \citenamefont
  {Bierhorst}, \citenamefont {Wayne}, \citenamefont {Stevens}, \citenamefont
  {Gerrits}, \citenamefont {Glancy}, \citenamefont {Hamel}, \citenamefont
  {Allman}, \citenamefont {Coakley}, \citenamefont {Dyer}, \citenamefont
  {Hodge}, \citenamefont {Lita}, \citenamefont {Verma}, \citenamefont
  {Lambrocco}, \citenamefont {Tortorici}, \citenamefont {Migdall},
  \citenamefont {Zhang}, \citenamefont {Kumor}, \citenamefont {Farr},
  \citenamefont {Marsili}, \citenamefont {Shaw}, \citenamefont {Stern},
  \citenamefont {Abellan}, \citenamefont {Amaya}, \citenamefont {Pruneri},
  \citenamefont {Jennewein}, \citenamefont {Mitchell}, \citenamefont {Kwiat},
  \citenamefont {Bienfang}, \citenamefont {Mirin}, \citenamefont {Knill},\ and\
  \citenamefont {Nam}}]{shalm_strong_2015}%
  \BibitemOpen
  \bibfield  {author} {\bibinfo {author} {\bibfnamefont {L.K.}\ \bibnamefont
  {Shalm}}, \bibinfo {author} {\bibfnamefont {E.}~\bibnamefont {Meyer-Scott}},
  \bibinfo {author} {\bibfnamefont {B.G.}\ \bibnamefont {Christensen}},
  \bibinfo {author} {\bibfnamefont {P.}~\bibnamefont {Bierhorst}}, \bibinfo
  {author} {\bibfnamefont {M.A.}\ \bibnamefont {Wayne}}, \bibinfo {author}
  {\bibfnamefont {M.J.}\ \bibnamefont {Stevens}}, \bibinfo {author}
  {\bibfnamefont {T.}~\bibnamefont {Gerrits}}, \bibinfo {author} {\bibfnamefont
  {S.}~\bibnamefont {Glancy}}, \bibinfo {author} {\bibfnamefont {D.R.}\
  \bibnamefont {Hamel}}, \bibinfo {author} {\bibfnamefont {M.S.}\ \bibnamefont
  {Allman}}, \bibinfo {author} {\bibfnamefont {K.J.}\ \bibnamefont {Coakley}},
  \bibinfo {author} {\bibfnamefont {S.D.}\ \bibnamefont {Dyer}}, \bibinfo
  {author} {\bibfnamefont {C.}~\bibnamefont {Hodge}}, \bibinfo {author}
  {\bibfnamefont {A.E.}\ \bibnamefont {Lita}}, \bibinfo {author} {\bibfnamefont
  {V.B.}\ \bibnamefont {Verma}}, \bibinfo {author} {\bibfnamefont
  {C.}~\bibnamefont {Lambrocco}}, \bibinfo {author} {\bibfnamefont
  {E.}~\bibnamefont {Tortorici}}, \bibinfo {author} {\bibfnamefont {A.L.}\
  \bibnamefont {Migdall}}, \bibinfo {author} {\bibfnamefont {Y.}~\bibnamefont
  {Zhang}}, \bibinfo {author} {\bibfnamefont {D.R.}\ \bibnamefont {Kumor}},
  \bibinfo {author} {\bibfnamefont {W.H.}\ \bibnamefont {Farr}}, \bibinfo
  {author} {\bibfnamefont {F.}~\bibnamefont {Marsili}}, \bibinfo {author}
  {\bibfnamefont {M.D.}\ \bibnamefont {Shaw}}, \bibinfo {author} {\bibfnamefont
  {J.A.}\ \bibnamefont {Stern}}, \bibinfo {author} {\bibfnamefont
  {C.}~\bibnamefont {Abellan}}, \bibinfo {author} {\bibfnamefont
  {W.}~\bibnamefont {Amaya}}, \bibinfo {author} {\bibfnamefont
  {V.}~\bibnamefont {Pruneri}}, \bibinfo {author} {\bibfnamefont
  {T.}~\bibnamefont {Jennewein}}, \bibinfo {author} {\bibfnamefont {M.W.}\
  \bibnamefont {Mitchell}}, \bibinfo {author} {\bibfnamefont {P.G.}\
  \bibnamefont {Kwiat}}, \bibinfo {author} {\bibfnamefont {J.C.}\ \bibnamefont
  {Bienfang}}, \bibinfo {author} {\bibfnamefont {R.P.}\ \bibnamefont {Mirin}},
  \bibinfo {author} {\bibfnamefont {E.}~\bibnamefont {Knill}}, \ and\ \bibinfo
  {author} {\bibfnamefont {S.W.}\ \bibnamefont {Nam}},\ }\bibfield  {title}
  {\enquote {\bibinfo {title} {Strong {Loophole}-{Free} {Test} of {Local}
  {Realism}},}\ }\href {\doibase 10.1103/PhysRevLett.115.250402} {\bibfield
  {journal} {\bibinfo  {journal} {Phys. Rev. Lett.}\ }\textbf {\bibinfo
  {volume} {115}},\ \bibinfo {pages} {250402} (\bibinfo {year}
  {2015})}\BibitemShut {NoStop}%
\bibitem [{\citenamefont {Wei}\ \emph {et~al.}(2005)\citenamefont {Wei},
  \citenamefont {Altepeter}, \citenamefont {Branning}, \citenamefont
  {Goldbart}, \citenamefont {James}, \citenamefont {Jeffrey}, \citenamefont
  {Kwiat}, \citenamefont {Mukhopadhyay},\ and\ \citenamefont
  {Peters}}]{wei_synthesizing_2005}%
  \BibitemOpen
  \bibfield  {author} {\bibinfo {author} {\bibfnamefont {Tzu-Chieh}\
  \bibnamefont {Wei}}, \bibinfo {author} {\bibfnamefont {Joseph~B.}\
  \bibnamefont {Altepeter}}, \bibinfo {author} {\bibfnamefont {David}\
  \bibnamefont {Branning}}, \bibinfo {author} {\bibfnamefont {Paul~M.}\
  \bibnamefont {Goldbart}}, \bibinfo {author} {\bibfnamefont {D.~F.~V.}\
  \bibnamefont {James}}, \bibinfo {author} {\bibfnamefont {Evan}\ \bibnamefont
  {Jeffrey}}, \bibinfo {author} {\bibfnamefont {Paul~G.}\ \bibnamefont
  {Kwiat}}, \bibinfo {author} {\bibfnamefont {Swagatam}\ \bibnamefont
  {Mukhopadhyay}}, \ and\ \bibinfo {author} {\bibfnamefont {Nicholas~A.}\
  \bibnamefont {Peters}},\ }\bibfield  {title} {\enquote {\bibinfo {title}
  {Synthesizing arbitrary two-photon polarization mixed states},}\ }\href
  {\doibase 10.1103/PhysRevA.71.032329} {\bibfield  {journal} {\bibinfo
  {journal} {Phys. Rev. A}\ }\textbf {\bibinfo {volume} {71}},\ \bibinfo
  {pages} {032329} (\bibinfo {year} {2005})}\BibitemShut {NoStop}%
\bibitem [{\citenamefont {Brendel}\ \emph {et~al.}(1999)\citenamefont
  {Brendel}, \citenamefont {Gisin}, \citenamefont {Tittel},\ and\ \citenamefont
  {Zbinden}}]{brendel_pulsed_1999}%
  \BibitemOpen
  \bibfield  {author} {\bibinfo {author} {\bibfnamefont {J.}~\bibnamefont
  {Brendel}}, \bibinfo {author} {\bibfnamefont {N.}~\bibnamefont {Gisin}},
  \bibinfo {author} {\bibfnamefont {W.}~\bibnamefont {Tittel}}, \ and\ \bibinfo
  {author} {\bibfnamefont {H.}~\bibnamefont {Zbinden}},\ }\bibfield  {title}
  {\enquote {\bibinfo {title} {Pulsed {Energy}-{Time} {Entangled}
  {Twin}-{Photon} {Source} for {Quantum} {Communication}},}\ }\href {\doibase
  10.1103/PhysRevLett.82.2594} {\bibfield  {journal} {\bibinfo  {journal}
  {Phys. Rev. Lett.}\ }\textbf {\bibinfo {volume} {82}},\ \bibinfo {pages}
  {2594--2597} (\bibinfo {year} {1999})}\BibitemShut {NoStop}%
\bibitem [{\citenamefont {Mair}\ \emph {et~al.}(2001)\citenamefont {Mair},
  \citenamefont {Vaziri}, \citenamefont {Weihs},\ and\ \citenamefont
  {Zeilinger}}]{mair_entanglement_2001}%
  \BibitemOpen
  \bibfield  {author} {\bibinfo {author} {\bibfnamefont {Alois}\ \bibnamefont
  {Mair}}, \bibinfo {author} {\bibfnamefont {Alipasha}\ \bibnamefont {Vaziri}},
  \bibinfo {author} {\bibfnamefont {Gregor}\ \bibnamefont {Weihs}}, \ and\
  \bibinfo {author} {\bibfnamefont {Anton}\ \bibnamefont {Zeilinger}},\
  }\bibfield  {title} {\enquote {\bibinfo {title} {Entanglement of the orbital
  angular momentum states of photons},}\ }\href {\doibase 10.1038/35085529}
  {\bibfield  {journal} {\bibinfo  {journal} {Nature}\ }\textbf {\bibinfo
  {volume} {412}},\ \bibinfo {pages} {313--316} (\bibinfo {year}
  {2001})}\BibitemShut {NoStop}%
\bibitem [{\citenamefont {Kwiat}\ \emph {et~al.}(1995)\citenamefont {Kwiat},
  \citenamefont {Mattle}, \citenamefont {Weinfurter}, \citenamefont
  {Zeilinger}, \citenamefont {Sergienko},\ and\ \citenamefont
  {Shih}}]{kwiat_new_1995}%
  \BibitemOpen
  \bibfield  {author} {\bibinfo {author} {\bibfnamefont {Paul~G.}\ \bibnamefont
  {Kwiat}}, \bibinfo {author} {\bibfnamefont {Klaus}\ \bibnamefont {Mattle}},
  \bibinfo {author} {\bibfnamefont {Harald}\ \bibnamefont {Weinfurter}},
  \bibinfo {author} {\bibfnamefont {Anton}\ \bibnamefont {Zeilinger}}, \bibinfo
  {author} {\bibfnamefont {Alexander~V.}\ \bibnamefont {Sergienko}}, \ and\
  \bibinfo {author} {\bibfnamefont {Yanhua}\ \bibnamefont {Shih}},\ }\bibfield
  {title} {\enquote {\bibinfo {title} {New {High}-{Intensity} {Source} of
  {Polarization}-{Entangled} {Photon} {Pairs}},}\ }\href {\doibase
  10.1103/PhysRevLett.75.4337} {\bibfield  {journal} {\bibinfo  {journal}
  {Phys. Rev. Lett.}\ }\textbf {\bibinfo {volume} {75}},\ \bibinfo {pages}
  {4337--4341} (\bibinfo {year} {1995})}\BibitemShut {NoStop}%
\bibitem [{\citenamefont {Thew}\ \emph {et~al.}(2002)\citenamefont {Thew},
  \citenamefont {Tanzilli}, \citenamefont {Tittel}, \citenamefont {Zbinden},\
  and\ \citenamefont {Gisin}}]{thew_experimental_2002}%
  \BibitemOpen
  \bibfield  {author} {\bibinfo {author} {\bibfnamefont {R.~T.}\ \bibnamefont
  {Thew}}, \bibinfo {author} {\bibfnamefont {S.}~\bibnamefont {Tanzilli}},
  \bibinfo {author} {\bibfnamefont {W.}~\bibnamefont {Tittel}}, \bibinfo
  {author} {\bibfnamefont {H.}~\bibnamefont {Zbinden}}, \ and\ \bibinfo
  {author} {\bibfnamefont {N.}~\bibnamefont {Gisin}},\ }\bibfield  {title}
  {\enquote {\bibinfo {title} {Experimental investigation of the robustness of
  partially entangled qubits over 11 km},}\ }\href {\doibase
  10.1103/PhysRevA.66.062304} {\bibfield  {journal} {\bibinfo  {journal} {Phys.
  Rev. A}\ }\textbf {\bibinfo {volume} {66}},\ \bibinfo {pages} {062304}
  (\bibinfo {year} {2002})}\BibitemShut {NoStop}%
\bibitem [{\citenamefont {Bernhard}\ \emph {et~al.}(2013)\citenamefont
  {Bernhard}, \citenamefont {Bessire}, \citenamefont {Feurer},\ and\
  \citenamefont {Stefanov}}]{bernhard_shaping_2013}%
  \BibitemOpen
  \bibfield  {author} {\bibinfo {author} {\bibfnamefont {Christof}\
  \bibnamefont {Bernhard}}, \bibinfo {author} {\bibfnamefont {Bänz}\
  \bibnamefont {Bessire}}, \bibinfo {author} {\bibfnamefont {Thomas}\
  \bibnamefont {Feurer}}, \ and\ \bibinfo {author} {\bibfnamefont {André}\
  \bibnamefont {Stefanov}},\ }\bibfield  {title} {\enquote {\bibinfo {title}
  {Shaping frequency-entangled qudits},}\ }\href {\doibase
  10.1103/PhysRevA.88.032322} {\bibfield  {journal} {\bibinfo  {journal} {Phys.
  Rev. A}\ }\textbf {\bibinfo {volume} {88}},\ \bibinfo {pages} {032322}
  (\bibinfo {year} {2013})}\BibitemShut {NoStop}%
\bibitem [{\citenamefont {Kwiat}\ \emph {et~al.}(1999)\citenamefont {Kwiat},
  \citenamefont {Waks}, \citenamefont {White}, \citenamefont {Appelbaum},\ and\
  \citenamefont {Eberhard}}]{kwiat_ultrabright_1999}%
  \BibitemOpen
  \bibfield  {author} {\bibinfo {author} {\bibfnamefont {Paul~G.}\ \bibnamefont
  {Kwiat}}, \bibinfo {author} {\bibfnamefont {Edo}\ \bibnamefont {Waks}},
  \bibinfo {author} {\bibfnamefont {Andrew~G.}\ \bibnamefont {White}}, \bibinfo
  {author} {\bibfnamefont {Ian}\ \bibnamefont {Appelbaum}}, \ and\ \bibinfo
  {author} {\bibfnamefont {Philippe~H.}\ \bibnamefont {Eberhard}},\ }\bibfield
  {title} {\enquote {\bibinfo {title} {Ultrabright source of
  polarization-entangled photons},}\ }\href {\doibase 10.1103/PhysRevA.60.R773}
  {\bibfield  {journal} {\bibinfo  {journal} {Phys. Rev. A}\ }\textbf {\bibinfo
  {volume} {60}},\ \bibinfo {pages} {R773--R776} (\bibinfo {year}
  {1999})}\BibitemShut {NoStop}%
\bibitem [{\citenamefont {Mattle}\ \emph {et~al.}(1996)\citenamefont {Mattle},
  \citenamefont {Weinfurter}, \citenamefont {Kwiat},\ and\ \citenamefont
  {Zeilinger}}]{mattle_dense_1996}%
  \BibitemOpen
  \bibfield  {author} {\bibinfo {author} {\bibfnamefont {Klaus}\ \bibnamefont
  {Mattle}}, \bibinfo {author} {\bibfnamefont {Harald}\ \bibnamefont
  {Weinfurter}}, \bibinfo {author} {\bibfnamefont {Paul~G.}\ \bibnamefont
  {Kwiat}}, \ and\ \bibinfo {author} {\bibfnamefont {Anton}\ \bibnamefont
  {Zeilinger}},\ }\bibfield  {title} {\enquote {\bibinfo {title} {Dense
  {Coding} in {Experimental} {Quantum} {Communication}},}\ }\href {\doibase
  10.1103/PhysRevLett.76.4656} {\bibfield  {journal} {\bibinfo  {journal}
  {Phys. Rev. Lett.}\ }\textbf {\bibinfo {volume} {76}},\ \bibinfo {pages}
  {4656--4659} (\bibinfo {year} {1996})}\BibitemShut {NoStop}%
\bibitem [{\citenamefont {Wu}\ \emph {et~al.}(2019)\citenamefont {Wu},
  \citenamefont {Liu}, \citenamefont {Chen},\ and\ \citenamefont
  {Chuu}}]{wu_revival_2019}%
  \BibitemOpen
  \bibfield  {author} {\bibinfo {author} {\bibfnamefont {Chih-Hsiang}\
  \bibnamefont {Wu}}, \bibinfo {author} {\bibfnamefont {Chiao-Kai}\
  \bibnamefont {Liu}}, \bibinfo {author} {\bibfnamefont {Yi-Cheng}\
  \bibnamefont {Chen}}, \ and\ \bibinfo {author} {\bibfnamefont {Chih-Sung}\
  \bibnamefont {Chuu}},\ }\bibfield  {title} {\enquote {\bibinfo {title}
  {Revival of {Quantum} {Interference} by {Modulating} the {Biphotons}},}\
  }\href {\doibase 10.1103/PhysRevLett.123.143601} {\bibfield  {journal}
  {\bibinfo  {journal} {Phys. Rev. Lett.}\ }\textbf {\bibinfo {volume} {123}},\
  \bibinfo {pages} {143601} (\bibinfo {year} {2019})}\BibitemShut {NoStop}%
\bibitem [{\citenamefont {Barreiro}\ \emph {et~al.}(2005)\citenamefont
  {Barreiro}, \citenamefont {Langford}, \citenamefont {Peters},\ and\
  \citenamefont {Kwiat}}]{barreiro_generation_2005}%
  \BibitemOpen
  \bibfield  {author} {\bibinfo {author} {\bibfnamefont {Julio~T.}\
  \bibnamefont {Barreiro}}, \bibinfo {author} {\bibfnamefont {Nathan~K.}\
  \bibnamefont {Langford}}, \bibinfo {author} {\bibfnamefont {Nicholas~A.}\
  \bibnamefont {Peters}}, \ and\ \bibinfo {author} {\bibfnamefont {Paul~G.}\
  \bibnamefont {Kwiat}},\ }\bibfield  {title} {\enquote {\bibinfo {title}
  {Generation of {Hyperentangled} {Photon} {Pairs}},}\ }\href {\doibase
  10.1103/PhysRevLett.95.260501} {\bibfield  {journal} {\bibinfo  {journal}
  {Phys. Rev. Lett.}\ }\textbf {\bibinfo {volume} {95}},\ \bibinfo {pages}
  {260501} (\bibinfo {year} {2005})}\BibitemShut {NoStop}%
\bibitem [{\citenamefont {Chen}\ \emph {et~al.}(2020)\citenamefont {Chen},
  \citenamefont {Riazi}, \citenamefont {Zhu},\ and\ \citenamefont
  {Qian}}]{chen_recovering_2020}%
  \BibitemOpen
  \bibfield  {author} {\bibinfo {author} {\bibfnamefont {Changjia}\
  \bibnamefont {Chen}}, \bibinfo {author} {\bibfnamefont {Arash}\ \bibnamefont
  {Riazi}}, \bibinfo {author} {\bibfnamefont {Eric~Y.}\ \bibnamefont {Zhu}}, \
  and\ \bibinfo {author} {\bibfnamefont {Li}~\bibnamefont {Qian}},\ }\bibfield
  {title} {\enquote {\bibinfo {title} {Recovering the full dimensionality of
  hyperentanglement in collinear photon pairs},}\ }\href {\doibase
  10.1103/PhysRevA.101.013834} {\bibfield  {journal} {\bibinfo  {journal}
  {Phys. Rev. A}\ }\textbf {\bibinfo {volume} {101}},\ \bibinfo {pages}
  {013834} (\bibinfo {year} {2020})}\BibitemShut {NoStop}%
\bibitem [{\citenamefont {Guo}\ \emph {et~al.}(2015)\citenamefont {Guo},
  \citenamefont {Chen}, \citenamefont {Shu}, \citenamefont {Loy},\ and\
  \citenamefont {Du}}]{guo_frequency-bin_2015}%
  \BibitemOpen
  \bibfield  {author} {\bibinfo {author} {\bibfnamefont {Xianxin}\ \bibnamefont
  {Guo}}, \bibinfo {author} {\bibfnamefont {Peng}\ \bibnamefont {Chen}},
  \bibinfo {author} {\bibfnamefont {Chi}\ \bibnamefont {Shu}}, \bibinfo
  {author} {\bibfnamefont {M.~M.~T.}\ \bibnamefont {Loy}}, \ and\ \bibinfo
  {author} {\bibfnamefont {Shengwang}\ \bibnamefont {Du}},\ }\bibfield  {title}
  {\enquote {\bibinfo {title} {Frequency-bin entanglement with tunable
  phase},}\ }\href {\doibase 10.1088/2040-8978/17/10/105201} {\bibfield
  {journal} {\bibinfo  {journal} {J. Opt.}\ }\textbf {\bibinfo {volume} {17}},\
  \bibinfo {pages} {105201} (\bibinfo {year} {2015})}\BibitemShut {NoStop}%
\bibitem [{\citenamefont {Shu}\ \emph {et~al.}(2015)\citenamefont {Shu},
  \citenamefont {Guo}, \citenamefont {Chen}, \citenamefont {Loy},\ and\
  \citenamefont {Du}}]{shu_narrowband_2015}%
  \BibitemOpen
  \bibfield  {author} {\bibinfo {author} {\bibfnamefont {Chi}\ \bibnamefont
  {Shu}}, \bibinfo {author} {\bibfnamefont {Xianxin}\ \bibnamefont {Guo}},
  \bibinfo {author} {\bibfnamefont {Peng}\ \bibnamefont {Chen}}, \bibinfo
  {author} {\bibfnamefont {M.~M.~T.}\ \bibnamefont {Loy}}, \ and\ \bibinfo
  {author} {\bibfnamefont {Shengwang}\ \bibnamefont {Du}},\ }\bibfield  {title}
  {\enquote {\bibinfo {title} {Narrowband biphotons with
  polarization-frequency-coupled entanglement},}\ }\href {\doibase
  10.1103/PhysRevA.91.043820} {\bibfield  {journal} {\bibinfo  {journal} {Phys.
  Rev. A}\ }\textbf {\bibinfo {volume} {91}},\ \bibinfo {pages} {043820}
  (\bibinfo {year} {2015})}\BibitemShut {NoStop}%
\bibitem [{\citenamefont {Chen}\ \emph {et~al.}(2015)\citenamefont {Chen},
  \citenamefont {Guo}, \citenamefont {Shu}, \citenamefont {Loy},\ and\
  \citenamefont {Du}}]{chen_frequency-induced_2015}%
  \BibitemOpen
  \bibfield  {author} {\bibinfo {author} {\bibfnamefont {Peng}\ \bibnamefont
  {Chen}}, \bibinfo {author} {\bibfnamefont {Xianxin}\ \bibnamefont {Guo}},
  \bibinfo {author} {\bibfnamefont {Chi}\ \bibnamefont {Shu}}, \bibinfo
  {author} {\bibfnamefont {M.~M.~T.}\ \bibnamefont {Loy}}, \ and\ \bibinfo
  {author} {\bibfnamefont {Shengwang}\ \bibnamefont {Du}},\ }\bibfield  {title}
  {\enquote {\bibinfo {title} {Frequency-induced phase-tunable
  polarization-entangled narrowband biphotons},}\ }\href {\doibase
  10.1364/OPTICA.2.000505} {\bibfield  {journal} {\bibinfo  {journal} {Optica}\
  }\textbf {\bibinfo {volume} {2}},\ \bibinfo {pages} {505--508} (\bibinfo
  {year} {2015})}\BibitemShut {NoStop}%
\bibitem [{\citenamefont {Ciampini}\ \emph {et~al.}(2016)\citenamefont
  {Ciampini}, \citenamefont {Orieux}, \citenamefont {Paesani}, \citenamefont
  {Sciarrino}, \citenamefont {Corrielli}, \citenamefont {Crespi}, \citenamefont
  {Ramponi}, \citenamefont {Osellame},\ and\ \citenamefont
  {Mataloni}}]{ciampini_path-polarization_2016}%
  \BibitemOpen
  \bibfield  {author} {\bibinfo {author} {\bibfnamefont {Mario~Arnolfo}\
  \bibnamefont {Ciampini}}, \bibinfo {author} {\bibfnamefont {Adeline}\
  \bibnamefont {Orieux}}, \bibinfo {author} {\bibfnamefont {Stefano}\
  \bibnamefont {Paesani}}, \bibinfo {author} {\bibfnamefont {Fabio}\
  \bibnamefont {Sciarrino}}, \bibinfo {author} {\bibfnamefont {Giacomo}\
  \bibnamefont {Corrielli}}, \bibinfo {author} {\bibfnamefont {Andrea}\
  \bibnamefont {Crespi}}, \bibinfo {author} {\bibfnamefont {Roberta}\
  \bibnamefont {Ramponi}}, \bibinfo {author} {\bibfnamefont {Roberto}\
  \bibnamefont {Osellame}}, \ and\ \bibinfo {author} {\bibfnamefont {Paolo}\
  \bibnamefont {Mataloni}},\ }\bibfield  {title} {\enquote {\bibinfo {title}
  {Path-polarization hyperentangled and cluster states of photons on a chip},}\
  }\href@noop {} {\bibfield  {journal} {\bibinfo  {journal} {Light-Sci. Appl.}\
  }\textbf {\bibinfo {volume} {5}},\ \bibinfo {pages} {e16064--e16064}
  (\bibinfo {year} {2016})}\BibitemShut {NoStop}%
\bibitem [{\citenamefont {Reimer}\ \emph {et~al.}(2019)\citenamefont {Reimer},
  \citenamefont {Sciara}, \citenamefont {Roztocki}, \citenamefont {Islam},
  \citenamefont {Romero~Cortés}, \citenamefont {Zhang}, \citenamefont
  {Fischer}, \citenamefont {Loranger}, \citenamefont {Kashyap}, \citenamefont
  {Cino}, \citenamefont {Chu}, \citenamefont {Little}, \citenamefont {Moss},
  \citenamefont {Caspani}, \citenamefont {Munro}, \citenamefont {Azaña},
  \citenamefont {Kues},\ and\ \citenamefont
  {Morandotti}}]{reimer_high-dimensional_2019}%
  \BibitemOpen
  \bibfield  {author} {\bibinfo {author} {\bibfnamefont {Christian}\
  \bibnamefont {Reimer}}, \bibinfo {author} {\bibfnamefont {Stefania}\
  \bibnamefont {Sciara}}, \bibinfo {author} {\bibfnamefont {Piotr}\
  \bibnamefont {Roztocki}}, \bibinfo {author} {\bibfnamefont {Mehedi}\
  \bibnamefont {Islam}}, \bibinfo {author} {\bibfnamefont {Luis}\ \bibnamefont
  {Romero~Cortés}}, \bibinfo {author} {\bibfnamefont {Yanbing}\ \bibnamefont
  {Zhang}}, \bibinfo {author} {\bibfnamefont {Bennet}\ \bibnamefont {Fischer}},
  \bibinfo {author} {\bibfnamefont {Sébastien}\ \bibnamefont {Loranger}},
  \bibinfo {author} {\bibfnamefont {Raman}\ \bibnamefont {Kashyap}}, \bibinfo
  {author} {\bibfnamefont {Alfonso}\ \bibnamefont {Cino}}, \bibinfo {author}
  {\bibfnamefont {Sai~T.}\ \bibnamefont {Chu}}, \bibinfo {author}
  {\bibfnamefont {Brent~E.}\ \bibnamefont {Little}}, \bibinfo {author}
  {\bibfnamefont {David~J.}\ \bibnamefont {Moss}}, \bibinfo {author}
  {\bibfnamefont {Lucia}\ \bibnamefont {Caspani}}, \bibinfo {author}
  {\bibfnamefont {William~J.}\ \bibnamefont {Munro}}, \bibinfo {author}
  {\bibfnamefont {José}\ \bibnamefont {Azaña}}, \bibinfo {author}
  {\bibfnamefont {Michael}\ \bibnamefont {Kues}}, \ and\ \bibinfo {author}
  {\bibfnamefont {Roberto}\ \bibnamefont {Morandotti}},\ }\bibfield  {title}
  {\enquote {\bibinfo {title} {High-dimensional one-way quantum processing
  implemented on d-level cluster states},}\ }\href@noop {} {\bibfield
  {journal} {\bibinfo  {journal} {Nat. Phys.}\ }\textbf {\bibinfo {volume}
  {15}},\ \bibinfo {pages} {148--153} (\bibinfo {year} {2019})}\BibitemShut
  {NoStop}%
\bibitem [{\citenamefont {Spreeuw}(1998)}]{spreeuw_classical_1998}%
  \BibitemOpen
  \bibfield  {author} {\bibinfo {author} {\bibfnamefont {Robert J.~C.}\
  \bibnamefont {Spreeuw}},\ }\bibfield  {title} {\enquote {\bibinfo {title} {A
  {Classical} {Analogy} of {Entanglement}},}\ }\href@noop {} {\bibfield
  {journal} {\bibinfo  {journal} {Found. Phys.}\ }\textbf {\bibinfo {volume}
  {28}},\ \bibinfo {pages} {361--374} (\bibinfo {year} {1998})}\BibitemShut
  {NoStop}%
\bibitem [{\citenamefont {Karimi}\ and\ \citenamefont
  {Boyd}(2015)}]{karimi_classical_2015}%
  \BibitemOpen
  \bibfield  {author} {\bibinfo {author} {\bibfnamefont {Ebrahim}\ \bibnamefont
  {Karimi}}\ and\ \bibinfo {author} {\bibfnamefont {Robert~W.}\ \bibnamefont
  {Boyd}},\ }\bibfield  {title} {\enquote {\bibinfo {title} {Classical
  entanglement?}}\ }\href@noop {} {\bibfield  {journal} {\bibinfo  {journal}
  {Science}\ }\textbf {\bibinfo {volume} {350}},\ \bibinfo {pages} {1172--1173}
  (\bibinfo {year} {2015})}\BibitemShut {NoStop}%
\bibitem [{\citenamefont {Shen}\ and\ \citenamefont
  {Rosales-Guzmán}(2022)}]{shen_nonseparable_2022}%
  \BibitemOpen
  \bibfield  {author} {\bibinfo {author} {\bibfnamefont {Yijie}\ \bibnamefont
  {Shen}}\ and\ \bibinfo {author} {\bibfnamefont {Carmelo}\ \bibnamefont
  {Rosales-Guzmán}},\ }\bibfield  {title} {\enquote {\bibinfo {title}
  {Nonseparable {States} of {Light}: {From} {Quantum} to {Classical}},}\
  }\href@noop {} {\bibfield  {journal} {\bibinfo  {journal} {Laser \& Photonics
  Rev.}\ }\textbf {\bibinfo {volume} {16}},\ \bibinfo {pages} {2100533}
  (\bibinfo {year} {2022})}\BibitemShut {NoStop}%
\bibitem [{\citenamefont {Shen}\ \emph {et~al.}(2021)\citenamefont {Shen},
  \citenamefont {Nape}, \citenamefont {Yang}, \citenamefont {Fu}, \citenamefont
  {Gong}, \citenamefont {Naidoo},\ and\ \citenamefont
  {Forbes}}]{shen_creation_2021}%
  \BibitemOpen
  \bibfield  {author} {\bibinfo {author} {\bibfnamefont {Yijie}\ \bibnamefont
  {Shen}}, \bibinfo {author} {\bibfnamefont {Isaac}\ \bibnamefont {Nape}},
  \bibinfo {author} {\bibfnamefont {Xilin}\ \bibnamefont {Yang}}, \bibinfo
  {author} {\bibfnamefont {Xing}\ \bibnamefont {Fu}}, \bibinfo {author}
  {\bibfnamefont {Mali}\ \bibnamefont {Gong}}, \bibinfo {author} {\bibfnamefont
  {Darryl}\ \bibnamefont {Naidoo}}, \ and\ \bibinfo {author} {\bibfnamefont
  {Andrew}\ \bibnamefont {Forbes}},\ }\bibfield  {title} {\enquote {\bibinfo
  {title} {Creation and control of high-dimensional multi-partite classically
  entangled light},}\ }\href@noop {} {\bibfield  {journal} {\bibinfo  {journal}
  {Light-Sci. Appl.}\ }\textbf {\bibinfo {volume} {10}},\ \bibinfo {pages} {50}
  (\bibinfo {year} {2021})}\BibitemShut {NoStop}%
\bibitem [{\citenamefont {Marrucci}\ \emph {et~al.}(2006)\citenamefont
  {Marrucci}, \citenamefont {Manzo},\ and\ \citenamefont
  {Paparo}}]{marrucci_optical_2006}%
  \BibitemOpen
  \bibfield  {author} {\bibinfo {author} {\bibfnamefont {L.}~\bibnamefont
  {Marrucci}}, \bibinfo {author} {\bibfnamefont {C.}~\bibnamefont {Manzo}}, \
  and\ \bibinfo {author} {\bibfnamefont {D.}~\bibnamefont {Paparo}},\
  }\bibfield  {title} {\enquote {\bibinfo {title} {Optical {Spin}-to-{Orbital}
  {Angular} {Momentum} {Conversion} in {Inhomogeneous} {Anisotropic}
  {Media}},}\ }\href@noop {} {\bibfield  {journal} {\bibinfo  {journal} {Phys.
  Rev. Lett.}\ }\textbf {\bibinfo {volume} {96}},\ \bibinfo {pages} {163905}
  (\bibinfo {year} {2006})}\BibitemShut {NoStop}%
\bibitem [{\citenamefont {Devlin}\ \emph {et~al.}(2017)\citenamefont {Devlin},
  \citenamefont {Ambrosio}, \citenamefont {Rubin}, \citenamefont {Mueller},\
  and\ \citenamefont {Capasso}}]{devlin_arbitrary_2017}%
  \BibitemOpen
  \bibfield  {author} {\bibinfo {author} {\bibfnamefont {Robert~C.}\
  \bibnamefont {Devlin}}, \bibinfo {author} {\bibfnamefont {Antonio}\
  \bibnamefont {Ambrosio}}, \bibinfo {author} {\bibfnamefont {Noah~A.}\
  \bibnamefont {Rubin}}, \bibinfo {author} {\bibfnamefont {J.~P.~Balthasar}\
  \bibnamefont {Mueller}}, \ and\ \bibinfo {author} {\bibfnamefont {Federico}\
  \bibnamefont {Capasso}},\ }\bibfield  {title} {\enquote {\bibinfo {title}
  {Arbitrary spin-to–orbital angular momentum conversion of light},}\
  }\href@noop {} {\bibfield  {journal} {\bibinfo  {journal} {Science}\ }\textbf
  {\bibinfo {volume} {358}},\ \bibinfo {pages} {896--901} (\bibinfo {year}
  {2017})}\BibitemShut {NoStop}%
\bibitem [{\citenamefont {Wang}\ \emph {et~al.}(2019)\citenamefont {Wang},
  \citenamefont {Shi}, \citenamefont {Niu}, \citenamefont {Hua}, \citenamefont
  {Li}, \citenamefont {Zhu}, \citenamefont {Xie},\ and\ \citenamefont
  {Ye}}]{wang_multichannel_2019}%
  \BibitemOpen
  \bibfield  {author} {\bibinfo {author} {\bibfnamefont {Enliang}\ \bibnamefont
  {Wang}}, \bibinfo {author} {\bibfnamefont {Lina}\ \bibnamefont {Shi}},
  \bibinfo {author} {\bibfnamefont {Jiebin}\ \bibnamefont {Niu}}, \bibinfo
  {author} {\bibfnamefont {Yilei}\ \bibnamefont {Hua}}, \bibinfo {author}
  {\bibfnamefont {Hailiang}\ \bibnamefont {Li}}, \bibinfo {author}
  {\bibfnamefont {Xiaoli}\ \bibnamefont {Zhu}}, \bibinfo {author}
  {\bibfnamefont {Changqing}\ \bibnamefont {Xie}}, \ and\ \bibinfo {author}
  {\bibfnamefont {Tianchun}\ \bibnamefont {Ye}},\ }\bibfield  {title} {\enquote
  {\bibinfo {title} {Multichannel {Spatially} {Nonhomogeneous} {Focused}
  {Vector} {Vortex} {Beams} for {Quantum} {Experiments}},}\ }\href@noop {}
  {\bibfield  {journal} {\bibinfo  {journal} {Adv. Opt. Mater.}\ }\textbf
  {\bibinfo {volume} {7}},\ \bibinfo {pages} {1801415} (\bibinfo {year}
  {2019})}\BibitemShut {NoStop}%
\bibitem [{\citenamefont {Neves}\ \emph {et~al.}(2009)\citenamefont {Neves},
  \citenamefont {Lima}, \citenamefont {Delgado},\ and\ \citenamefont
  {Saavedra}}]{neves_hybrid_2009}%
  \BibitemOpen
  \bibfield  {author} {\bibinfo {author} {\bibfnamefont {Leonardo}\
  \bibnamefont {Neves}}, \bibinfo {author} {\bibfnamefont {Gustavo}\
  \bibnamefont {Lima}}, \bibinfo {author} {\bibfnamefont {Aldo}\ \bibnamefont
  {Delgado}}, \ and\ \bibinfo {author} {\bibfnamefont {Carlos}\ \bibnamefont
  {Saavedra}},\ }\bibfield  {title} {\enquote {\bibinfo {title} {Hybrid
  photonic entanglement: {Realization}, characterization, and applications},}\
  }\href {\doibase 10.1103/PhysRevA.80.042322} {\bibfield  {journal} {\bibinfo
  {journal} {Phys. Rev. A}\ }\textbf {\bibinfo {volume} {80}},\ \bibinfo
  {pages} {042322} (\bibinfo {year} {2009})}\BibitemShut {NoStop}%
\bibitem [{\citenamefont {Ma}\ \emph {et~al.}(2009)\citenamefont {Ma},
  \citenamefont {Qarry}, \citenamefont {Kofler}, \citenamefont {Jennewein},\
  and\ \citenamefont {Zeilinger}}]{ma_experimental_2009}%
  \BibitemOpen
  \bibfield  {author} {\bibinfo {author} {\bibfnamefont {Xiao-song}\
  \bibnamefont {Ma}}, \bibinfo {author} {\bibfnamefont {Angie}\ \bibnamefont
  {Qarry}}, \bibinfo {author} {\bibfnamefont {Johannes}\ \bibnamefont
  {Kofler}}, \bibinfo {author} {\bibfnamefont {Thomas}\ \bibnamefont
  {Jennewein}}, \ and\ \bibinfo {author} {\bibfnamefont {Anton}\ \bibnamefont
  {Zeilinger}},\ }\bibfield  {title} {\enquote {\bibinfo {title} {Experimental
  violation of a {Bell} inequality with two different degrees of freedom of
  entangled particle pairs},}\ }\href {\doibase 10.1103/PhysRevA.79.042101}
  {\bibfield  {journal} {\bibinfo  {journal} {Phys. Rev. A}\ }\textbf {\bibinfo
  {volume} {79}},\ \bibinfo {pages} {042101} (\bibinfo {year}
  {2009})}\BibitemShut {NoStop}%
\bibitem [{\citenamefont {Gabriel}\ \emph {et~al.}(2011)\citenamefont
  {Gabriel}, \citenamefont {Aiello}, \citenamefont {Zhong}, \citenamefont
  {Euser}, \citenamefont {Joly}, \citenamefont {Banzer}, \citenamefont
  {F\"{o}rtsch}, \citenamefont {Elser}, \citenamefont {Andersen}, \citenamefont
  {Marquardt}, \citenamefont {Russell},\ and\ \citenamefont
  {Leuchs}}]{gabriel_entangling_2011}%
  \BibitemOpen
  \bibfield  {author} {\bibinfo {author} {\bibfnamefont {C.}~\bibnamefont
  {Gabriel}}, \bibinfo {author} {\bibfnamefont {A.}~\bibnamefont {Aiello}},
  \bibinfo {author} {\bibfnamefont {W.}~\bibnamefont {Zhong}}, \bibinfo
  {author} {\bibfnamefont {T.~G.}\ \bibnamefont {Euser}}, \bibinfo {author}
  {\bibfnamefont {N.~Y.}\ \bibnamefont {Joly}}, \bibinfo {author}
  {\bibfnamefont {P.}~\bibnamefont {Banzer}}, \bibinfo {author} {\bibfnamefont
  {M.}~\bibnamefont {F\"{o}rtsch}}, \bibinfo {author} {\bibfnamefont
  {D.}~\bibnamefont {Elser}}, \bibinfo {author} {\bibfnamefont {U.~L.}\
  \bibnamefont {Andersen}}, \bibinfo {author} {\bibfnamefont {Ch.}\
  \bibnamefont {Marquardt}}, \bibinfo {author} {\bibfnamefont {P.~St.~J.}\
  \bibnamefont {Russell}}, \ and\ \bibinfo {author} {\bibfnamefont
  {G.}~\bibnamefont {Leuchs}},\ }\bibfield  {title} {\enquote {\bibinfo {title}
  {Entangling {Different} {Degrees} of {Freedom} by {Quadrature} {Squeezing}
  {Cylindrically} {Polarized} {Modes}},}\ }\href {\doibase
  10.1103/PhysRevLett.106.060502} {\bibfield  {journal} {\bibinfo  {journal}
  {Phys. Rev. Lett.}\ }\textbf {\bibinfo {volume} {106}},\ \bibinfo {pages}
  {060502} (\bibinfo {year} {2011})}\BibitemShut {NoStop}%
\bibitem [{\citenamefont {Wootters}(1998)}]{wootters_entanglement_1998}%
  \BibitemOpen
  \bibfield  {author} {\bibinfo {author} {\bibfnamefont {William~K.}\
  \bibnamefont {Wootters}},\ }\bibfield  {title} {\enquote {\bibinfo {title}
  {Entanglement of {Formation} of an {Arbitrary} {State} of {Two} {Qubits}},}\
  }\href {\doibase 10.1103/PhysRevLett.80.2245} {\bibfield  {journal} {\bibinfo
   {journal} {Phys. Rev. Lett.}\ }\textbf {\bibinfo {volume} {80}},\ \bibinfo
  {pages} {2245--2248} (\bibinfo {year} {1998})}\BibitemShut {NoStop}%
\bibitem [{\citenamefont {Riazi}\ \emph {et~al.}(2020)\citenamefont {Riazi},
  \citenamefont {Chen}, \citenamefont {Zhu}, \citenamefont {Gladyshev},
  \citenamefont {Kazansky}, \citenamefont {Sipe},\ and\ \citenamefont
  {Qian}}]{riazi_dispersion_2020}%
  \BibitemOpen
  \bibfield  {author} {\bibinfo {author} {\bibfnamefont {Arash}\ \bibnamefont
  {Riazi}}, \bibinfo {author} {\bibfnamefont {Changjia}\ \bibnamefont {Chen}},
  \bibinfo {author} {\bibfnamefont {Eric~Y.}\ \bibnamefont {Zhu}}, \bibinfo
  {author} {\bibfnamefont {Alexey~V.}\ \bibnamefont {Gladyshev}}, \bibinfo
  {author} {\bibfnamefont {Peter~G.}\ \bibnamefont {Kazansky}}, \bibinfo
  {author} {\bibfnamefont {J.~E.}\ \bibnamefont {Sipe}}, \ and\ \bibinfo
  {author} {\bibfnamefont {Li}~\bibnamefont {Qian}},\ }\bibfield  {title}
  {\enquote {\bibinfo {title} {Dispersion measurement assisted by a stimulated
  parametric process},}\ }\href {\doibase 10.1364/OL.387283} {\bibfield
  {journal} {\bibinfo  {journal} {Opt. Lett.}\ }\textbf {\bibinfo {volume}
  {45}},\ \bibinfo {pages} {2034--2037} (\bibinfo {year} {2020})}\BibitemShut
  {NoStop}%
\bibitem [{\citenamefont {Helt}\ \emph {et~al.}(2009)\citenamefont {Helt},
  \citenamefont {Zhu}, \citenamefont {Liscidini}, \citenamefont {Qian},\ and\
  \citenamefont {Sipe}}]{Helt_proposal_2009}%
  \BibitemOpen
  \bibfield  {author} {\bibinfo {author} {\bibfnamefont {L.~G.}\ \bibnamefont
  {Helt}}, \bibinfo {author} {\bibfnamefont {Eric~Y.}\ \bibnamefont {Zhu}},
  \bibinfo {author} {\bibfnamefont {Marco}\ \bibnamefont {Liscidini}}, \bibinfo
  {author} {\bibfnamefont {Li}~\bibnamefont {Qian}}, \ and\ \bibinfo {author}
  {\bibfnamefont {J.~E.}\ \bibnamefont {Sipe}},\ }\bibfield  {title} {\enquote
  {\bibinfo {title} {Proposal for in-fiber generation of telecom-band
  polarization-entangled photon pairs using a periodically poled fiber},}\
  }\href {\doibase 10.1364/OL.34.002138} {\bibfield  {journal} {\bibinfo
  {journal} {Opt. Lett.}\ }\textbf {\bibinfo {volume} {34}},\ \bibinfo {pages}
  {2138--2140} (\bibinfo {year} {2009})}\BibitemShut {NoStop}%
\bibitem [{\citenamefont {Zhu}\ \emph {et~al.}(2012)\citenamefont {Zhu},
  \citenamefont {Tang}, \citenamefont {Qian}, \citenamefont {Helt},
  \citenamefont {Liscidini}, \citenamefont {Sipe}, \citenamefont {Corbari},
  \citenamefont {Canagasabey}, \citenamefont {Ibsen},\ and\ \citenamefont
  {Kazansky}}]{zhu_direct_2012}%
  \BibitemOpen
  \bibfield  {author} {\bibinfo {author} {\bibfnamefont {Eric~Y.}\ \bibnamefont
  {Zhu}}, \bibinfo {author} {\bibfnamefont {Zhiyuan}\ \bibnamefont {Tang}},
  \bibinfo {author} {\bibfnamefont {Li}~\bibnamefont {Qian}}, \bibinfo {author}
  {\bibfnamefont {Lukas~G.}\ \bibnamefont {Helt}}, \bibinfo {author}
  {\bibfnamefont {Marco}\ \bibnamefont {Liscidini}}, \bibinfo {author}
  {\bibfnamefont {J.~E.}\ \bibnamefont {Sipe}}, \bibinfo {author}
  {\bibfnamefont {Costantino}\ \bibnamefont {Corbari}}, \bibinfo {author}
  {\bibfnamefont {Albert}\ \bibnamefont {Canagasabey}}, \bibinfo {author}
  {\bibfnamefont {Morten}\ \bibnamefont {Ibsen}}, \ and\ \bibinfo {author}
  {\bibfnamefont {Peter~G.}\ \bibnamefont {Kazansky}},\ }\bibfield  {title}
  {\enquote {\bibinfo {title} {Direct {Generation} of
  {Polarization}-{Entangled} {Photon} {Pairs} in a {Poled} {Fiber}},}\ }\href
  {\doibase 10.1103/PhysRevLett.108.213902} {\bibfield  {journal} {\bibinfo
  {journal} {Phys. Rev. Lett.}\ }\textbf {\bibinfo {volume} {108}},\ \bibinfo
  {pages} {213902} (\bibinfo {year} {2012})}\BibitemShut {NoStop}%
\bibitem [{\citenamefont {Chen}\ \emph {et~al.}(2017)\citenamefont {Chen},
  \citenamefont {Zhu}, \citenamefont {Riazi}, \citenamefont {Gladyshev},
  \citenamefont {Corbari}, \citenamefont {Ibsen}, \citenamefont {Kazansky},\
  and\ \citenamefont {Qian}}]{chen_compensation-free_2017}%
  \BibitemOpen
  \bibfield  {author} {\bibinfo {author} {\bibfnamefont {Changjia}\
  \bibnamefont {Chen}}, \bibinfo {author} {\bibfnamefont {Eric~Y.}\
  \bibnamefont {Zhu}}, \bibinfo {author} {\bibfnamefont {Arash}\ \bibnamefont
  {Riazi}}, \bibinfo {author} {\bibfnamefont {Alexey~V.}\ \bibnamefont
  {Gladyshev}}, \bibinfo {author} {\bibfnamefont {Costantino}\ \bibnamefont
  {Corbari}}, \bibinfo {author} {\bibfnamefont {Morten}\ \bibnamefont {Ibsen}},
  \bibinfo {author} {\bibfnamefont {Peter~G.}\ \bibnamefont {Kazansky}}, \ and\
  \bibinfo {author} {\bibfnamefont {Li}~\bibnamefont {Qian}},\ }\bibfield
  {title} {\enquote {\bibinfo {title} {Compensation-free broadband entangled
  photon pair sources},}\ }\href {\doibase 10.1364/OE.25.022667} {\bibfield
  {journal} {\bibinfo  {journal} {Opt. Express}\ }\textbf {\bibinfo {volume}
  {25}},\ \bibinfo {pages} {22667--22678} (\bibinfo {year} {2017})}\BibitemShut
  {NoStop}%
\bibitem [{\citenamefont {Chen}\ \emph {et~al.}(2018)\citenamefont {Chen},
  \citenamefont {Riazi}, \citenamefont {Zhu}, \citenamefont {Ng}, \citenamefont
  {Gladyshev}, \citenamefont {Kazansky},\ and\ \citenamefont
  {Qian}}]{chen_turn-key_2018}%
  \BibitemOpen
  \bibfield  {author} {\bibinfo {author} {\bibfnamefont {Changjia}\
  \bibnamefont {Chen}}, \bibinfo {author} {\bibfnamefont {Arash}\ \bibnamefont
  {Riazi}}, \bibinfo {author} {\bibfnamefont {Eric~Y.}\ \bibnamefont {Zhu}},
  \bibinfo {author} {\bibfnamefont {Mili}\ \bibnamefont {Ng}}, \bibinfo
  {author} {\bibfnamefont {Alexey~V.}\ \bibnamefont {Gladyshev}}, \bibinfo
  {author} {\bibfnamefont {Peter~G.}\ \bibnamefont {Kazansky}}, \ and\ \bibinfo
  {author} {\bibfnamefont {Li}~\bibnamefont {Qian}},\ }\bibfield  {title}
  {\enquote {\bibinfo {title} {Turn-key diode-pumped all-fiber broadband
  polarization-entangled photon source},}\ }\href {\doibase
  10.1364/OSAC.1.000981} {\bibfield  {journal} {\bibinfo  {journal} {OSA
  Continuum}\ }\textbf {\bibinfo {volume} {1}},\ \bibinfo {pages} {981--986}
  (\bibinfo {year} {2018})}\BibitemShut {NoStop}%
\bibitem [{\citenamefont {James}\ \emph {et~al.}(2001)\citenamefont {James},
  \citenamefont {Kwiat}, \citenamefont {Munro},\ and\ \citenamefont
  {White}}]{james_measurement_2001}%
  \BibitemOpen
  \bibfield  {author} {\bibinfo {author} {\bibfnamefont {Daniel F.~V.}\
  \bibnamefont {James}}, \bibinfo {author} {\bibfnamefont {Paul~G.}\
  \bibnamefont {Kwiat}}, \bibinfo {author} {\bibfnamefont {William~J.}\
  \bibnamefont {Munro}}, \ and\ \bibinfo {author} {\bibfnamefont {Andrew~G.}\
  \bibnamefont {White}},\ }\bibfield  {title} {\enquote {\bibinfo {title}
  {Measurement of qubits},}\ }\href {\doibase 10.1103/PhysRevA.64.052312}
  {\bibfield  {journal} {\bibinfo  {journal} {Phys. Rev. A}\ }\textbf {\bibinfo
  {volume} {64}},\ \bibinfo {pages} {052312} (\bibinfo {year}
  {2001})}\BibitemShut {NoStop}%
\end{thebibliography}%

\end{document}